\documentclass[%
preprint,
 amsmath,amssymb,
 aps,
]{revtex4-1}

\usepackage[utf8]{inputenc}
\usepackage[english]{babel}
\usepackage{ulem}

\usepackage{natbib}

\usepackage{graphicx}
\usepackage{dcolumn}
\usepackage{bm}
\usepackage{mathtools} 

\usepackage{subcaption}
\usepackage{hhline}

\newcommand{\bize}{\begin{itemize}}
\newcommand{\eize}{\end{itemize}}
\newcommand{\benum}{\begin{enumerate}}
\newcommand{\eenum}{\end{enumerate}}
\newcommand{\bdesc}{\begin{description}}
\newcommand{\edesc}{\end{description}}

\newcommand{\beqa}{\begin{eqnarray}}
\newcommand{\eeqa}{\end{eqnarray}}
\newcommand{\beq}{\begin{equation}}
\newcommand{\eeq}{\end{equation}}
\newcommand{\beqsub}{\begin{subequations}}
\newcommand{\eeqsub}{\end{subequations}}

\usepackage{color}
\definecolor{red}{rgb}{1,0,0} 
\definecolor{blue}{rgb}{0,0,0.8} 
\definecolor{green}{rgb}{0,0.5,0} 

\begin{document}

\preprint{APS/123-QED}

\title{Spatial Opinion Dynamics and the Effects of Two Types of Mixing}
\thanks{Support for this research comes from NIH grant P20GM104420 (BB and SK), the Computational Resources Core in the Institute for Bioinformatics and Evolutionary Studies under NIH grant P30GM103324, and the University of Idaho CLASS Excellence in Teaching the Humanities Endowment (BB).  RT was supported by NSERC (RGPIN-2016-05277).}%

\author{Bert O. Baumgaertner}
\email{bbaum@uidaho.edu}
 \affiliation{Department of Politics and Philosophy, University of Idaho.}

\author{Peter A. Fetros}%
 \email{fetr0509@vandals.uidaho.edu}
\affiliation{%
 Department of Electrical Engineering, University of Idaho
}%

\author{Stephen M. Krone}%
 \email{krone@uidaho.edu}
\affiliation{%
 Department of Mathematics, University of Idaho
}%

\author{Rebecca C. Tyson}
\email{rebecca.tyson@ubc.ca}
\affiliation{
Department of Mathematics, University of British Columbia, Okanagan
}%

\date{\today}

\begin{abstract}
Spatially-situated opinions that can be held with different degrees of conviction lead to spatio-temporal patterns such as clustering (homophily), polarization, and deadlock. Our goal is to understand how sensitive these patterns are to changes in the local nature of interactions. We introduce two different mixing mechanisms, spatial relocation and non-local interaction (``telephoning''), to an earlier fully spatial model (no mixing). Interestingly, the mechanisms that create deadlock in the fully spatial model have the opposite effect when there is a sufficient amount of mixing. With telephoning, not only is polarization and deadlock broken up, but consensus is hastened. The effects of mixing by relocation are even more pronounced.  Further insight into these dynamics is obtained for selected parameter regimes via comparison to the mean-field differential equations. 

\end{abstract}

\pacs{Valid PACS appear here}
\maketitle


\section{Introduction}

In \cite{baumgaertner2016opinion} we introduced and explored the behavior of a spatial model of opinion dynamics with an extended attitude spectrum in which opinions can become more or less entrenched. This {\it entrenchment model} allowed us to consider influence that is based on entrenchment or strength of opinion, as well as an echo chamber effect that occurs when like-minded individuals interact (also known as homophily). These mechanisms were found to promote clustering of like opinions and polarization toward more extreme attitudes, in turn creating deadlock. 

Attitudes in populations can be influenced through a variety of interaction types, some tending to occur locally and others via exchanges that are more wide-ranging. Here we ask how these forms of interaction affect patterns of clustering, polarization, and consensus of opinions. In particular, we add two types of `mixing' to the usual local influences.  These mixing mechanisms infuse local dynamics with non-local interactions. The first, which we call  {\it relocation}, involves individuals changing their physical locations. 
The second type of mixing, which we call {\it telephoning}, represents temporary interactions that people have with individuals outside of their usual `local' contacts, while retaining their spatial locations. Such interactions can occur, for example, during vacations, conferences, or community gatherings where individuals can have meaningful long-range interactions with  people with whom they do not regularly interact  (`meaningful' in the sense that the interaction has the potential to change an opinion).

Our result, in brief, is that the same mechanisms that cause deadlock in the fully spatial model have the opposite effect when there is a sufficient amount of mixing: consensus is reached rapidly. In our partially mixed spatial model, relocation disrupts spatial structure by moving individuals to new locations with some probability; telephoning maintains physical location but allows individuals to occasionally interact with individuals outside their `local' neighborhood. We specify the details of our fully spatial and partially mixed entrenchment models in Section \ref{model}. In Section \ref{wellmixedcase} we study the case of a well-mixed population (of infinite size) by considering the mean-field ordinary differential equation (ODE) model. The ODE generates expectations of how our spatial models will behave for sufficiently high levels of mixing. In Section \ref{results}, we compare the partially mixed models to the fully spatial model and to the ODE model. Finally, in Section \ref{discussion} we draw some comparisons with other opinion dynamics models. 

\section{The Spatial Model}
\label{model}

We begin by describing our agent-based discrete-time stochastic spatial model. The fully spatial version of this model (i.e., with no mixing) was introduced in \cite{baumgaertner2016opinion}.  Individuals reside at sites on a 2-dimensional grid that wraps in both directions (creating a torus), one individual per site (with default grid size $101 \times 101$, a population size of 10,201 individuals, but significantly larger populations are also explored). Each individual has an opinion that can be held with varying strengths. An individual's ``attitude" will contain both the strength of their opinion and the opinion itself (indicated by $+$ or $-$). Thus, each individual has an opinion (or attitude) from the \textit{attitude spectrum}
\[
{\cal A}=\{\pm 1, \pm 2,\ldots , \pm L\}.
\]
Given a particular attitude from $\cal A$, the ``opinion" is determined by the sign of the attitude, while the ``strength of the opinion" is determined by the absolute value of the attitude (this setup resembles the work of \cite{martins2013building}).

The updating of attitudes/opinions in the model depends on ``influence,'' ``amplification,'' and ``mixing.'' At each time step, all individuals consider adjusting their attitudes synchronously. The time step begins with some designated fraction (possibly 0) of the population relocating. Then, each individual chooses some other individual for an interaction. This choice is made either ``locally" or ``globally," with the selection possibly influenced by the states of the neighbors.  The first of these individuals, the ``focal'' individual, is the one considering a change in attitude, and this change is in response to the attitude of the second individual, the ``interaction partner." Since the updates are synchronous, all these choices and results are based on the spatial configuration of attitudes at the previous time step. The three models we consider have the following ingredients.

\bigskip\noindent \textit{Influence}. The strength of an individual's opinion can affect the likelihood that that individual will affect others. 
We account for this variable likelihood with an \textit{influence function} $I(a)$, $a\in {\cal A}$, that gives the influence exerted by an individual with attitude $a$. We consider five different influence functions: 
 \begin{alignat}{3}
& {\text{Quadratic}}: & \qquad &  I(a)=|a|^2, & & \\
 & {\text{Linear}}: & \qquad & I(a)=|a|,  & & \\
  & {\text{Uniform}}: & \qquad &  I(a)=1,  \\
  & {\text{Co-Linear}}: & \qquad & I(a)=L+1-|a|,  & & \\
 & {\text{Co-Quadratic}}: & \qquad &  I(a)=(L+1-|a|)^2.  & & 
 \end{alignat}
Individuals with strongly held opinions will have more influence under the linear and quadratic functions; the co-linear and co-quadratic functions give more influence to moderately held opinions; the uniform function gives everyone the same influence. We will sometimes refer to the linear and quadratic functions as \textit{extremist} influence functions and the co-linear and co-quadratic functions as \textit{centrist} influence functions.

\begin{figure}
\centering
	\begin{subfigure}[t]{0.48\textwidth}
	\includegraphics[width=\textwidth]{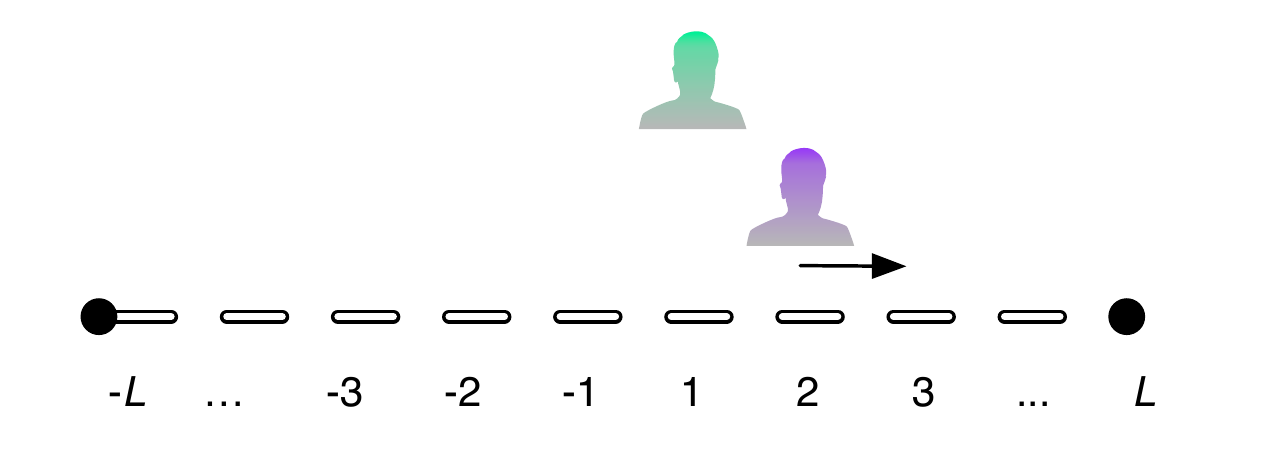}
	\caption{With Amplification}
    \end{subfigure}
    ~~
    \begin{subfigure}[t]{0.48\textwidth}
    \includegraphics[width=\textwidth]{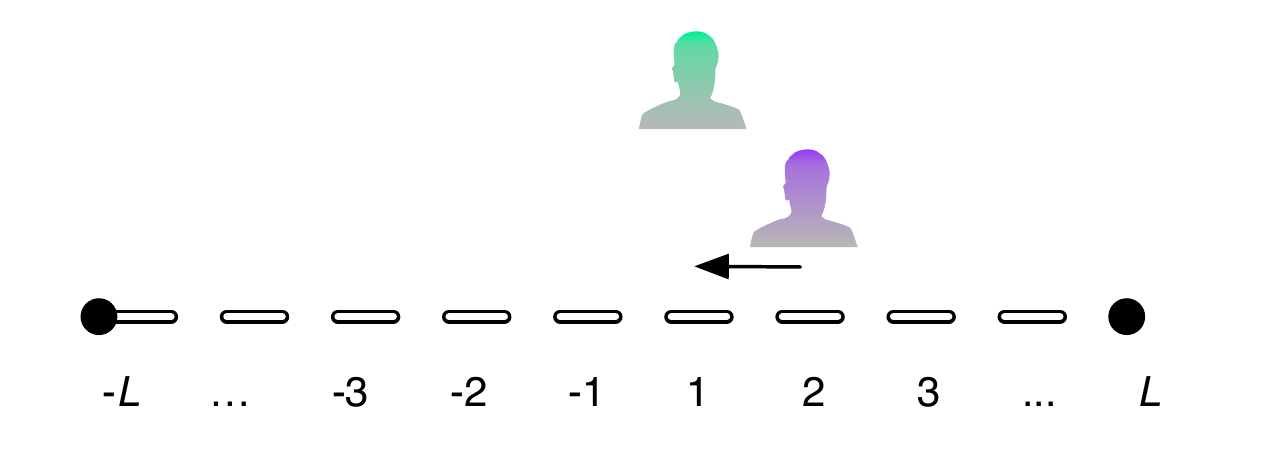}
    \caption{Without Ampflication}
    \end{subfigure}
\caption{\label{fig:amp_examples} (Color online) Examples of how a focal individual (purple at attitude 2) would update given an interaction with a like-minded individual (green at attitude 1). In an interaction with amplification (see (a)), the opinion of the focal individual becomes more entrenched, i.e., they adopt a more extreme attitude. In an interaction without amplification (see (b)), the focal individual changes their attitude towards the other individual's attitude.}
\end{figure}

\bigskip\noindent \textit{Amplification}.  When a ``focal'' individual looks to update its attitude via an interaction with another appropriately chosen individual, the result depends on whether the two opinions are on the same side of the attitude spectrum. If the attitudes are on opposite sides of the spectrum (i.e., the opinions are opposite), the focal individual will change its attitude by moving one step toward the other side. In the case where the opinions agree, two outcomes are possible: A fraction $p_a$ of the interactions result in a hardening of the opinion of the individual at the focal site $x$, 
while a fraction $1-p_a$ of these interactions result in no change (see Figure \ref{fig:amp_examples}). 
We refer to $p_a$ as the probability of ``opinion amplification.'' More formally, at a given time step, the attitude at focal site $x$, $A(x)$,  is updated following an interaction with the individual at site $z$ (appropriately chosen) according to one of these options as follows:

\medskip

{\it No opinion amplification}: With probability $1-p_a$, $A(x)$ is moved one allowable step toward the value of $A(z)$. Note that since there is no zero state in ${\cal A}$, a move to the left from $+1$ involves a jump to $-1$, and vice versa. If $A(z)=A(x)$ then $A(x)$ will not change.

{\it Opinion amplification}: With probability $p_a$, $A(x)$ is moved one (allowable) step to the right if $A(z)>0$ and one (allowable) step to the left if $A(z)<0$, regardless of where the value of $A(z)$ lies in relation to $A(x)$.   Clearly, the only possible movement for a maximally entrenched individual, i.e., $|A(x)|=L$, is toward the center.

Figures \ref{fig:amp_examples} \textcolor{blue}{and \ref{fig:amp_example2}} illustrate the difference between amplification and no amplification. Note that it is possible for amplification to produce echo chamber effects when individuals with the same opinion consistently interact (this is consistent with empirical findings, see  \cite{lord1979biased,miller1993attitude,munro2002biased,taber2006motivated}).

\begin{figure}
\begin{tabular}{cc} 
No amplification & Amplification \\[2ex]
\begin{tabular}{p{1.5cm}c}
& neighbor attitude $A(z)$ \\
focal attitude $A(x)$ &
\begin{tabular}{c||c|c|c|c}
 & -2 & -1 & 1 & 2 \\ \hhline{=||=|=|=|=|}
-2 & -2 & -$\mathbf{1}$ & -1 & -1 \\ \hline
-1 & -2 & -$\mathbf{1}$ & 1 & 1 \\ \hline
1 & -1 & -1 & $\mathbf{1}$ & 2 \\ \hline
2 & 1 & 1 & $\mathbf{1}$ & 2 \\ 
\end{tabular}
\end{tabular} & \begin{tabular}{p{1.5cm}c}
& neighbor attitude $A(z)$ \\
focal attitude $A(x)$ & 
\begin{tabular}{c||c|c|c|c}
 & -2 & -1 & 1 & 2 \\ \hhline{=||=|=|=|=|}
-2 & -2 & -$\mathbf{2}$ & -1 & -1 \\ \hline
-1 & -2 & -$\mathbf{2}$ & 1 & 1 \\ \hline
1 & -1 & -1 & $\mathbf{2}$ & 2 \\ \hline
2 & 1 & 1 & $\mathbf{2}$ & 2 \\ 
\end{tabular}
\end{tabular} 
\end{tabular}
\caption{Tables showing how a focal individual's attitude, $A(x)$, is updated given the selected neighbor's attitude, $A(z)$, where the attitude spectrum ranges from -2 to 2. The differences between no amplification and amplification are highlighted in boldface.
}
\label{fig:amp_example2}
\end{figure}

\bigskip\noindent \textit{Relocation}. Some fraction  $rel\in[0,1]$ of the population may be chosen to relocate by exchanging positions with other relocating individuals. More precisely, if the fraction $rel$ corresponds to $n$ individuals (i.e., $rel\  \times$ grid size $= n$, ), then $\lfloor n/2 \rfloor$ individuals are selected and each of these selects another individual at random to switch positions. This is carried out sequentially within a given time step. Some individuals may move more than once; if so, then fewer than $n$ individuals will have moved. Thus, $rel$ represents the maximum fraction of the population that relocates. In a given time step, any relocations will always take place before the interactions.

\bigskip\noindent \textit{Local and global interactions}.  When a focal individual interacts {\it locally}, it chooses one of its 8 nearest neighbors at random with probabilities weighted by the influences of these neighbors. If we denote the sites in the local neighborhood of $x$ as ${\cal N}(x)$ and if $A_t(y)$ is the attitude at site $y$ at the current time, then neighbor $z\in {\cal N}(x)$ is chosen with probability
\beq
\frac{I(A_t(z))}{\sum_{y\in {\cal N}(x)}I(A_t(y))}.
\label{eq:interaction-probability}
\eeq
When a focal individual interacts {\it globally}, it chooses one of the other individuals in the population at random, without regard to influence. (See below.) This includes a very small probability of choosing one of the 8 nearest neighbors. 

\bigskip

We now describe the three agent-based models, which we take to be variations on what we generally refer to as our entrenchment model of opinion dynamics. The first is the spatial model discussed in \cite{baumgaertner2016opinion}; the others introduce two forms of mixing into this model. We will analyze these spatial models in Section IV. Our goal is to compare the effects of the two different types of mixing, taken one at a time, to see how they differ from each other and from the fully spatial model. One of the points we wish to make is that there are several different types of ``mixing'' that one could consider, and they can produce different effects.

\begin{description}
\item [Fully Spatial Model]  All individuals interact locally and there is no relocation or telephoning.
\end{description}

\begin{description}
\item [Relocation Model]  All individuals interact locally but, prior to the interactions in each time step, a fraction $rel\in[0,1]$ of the population is randomly selected, two at a time, and the locations of the individuals are swapped. 
\end{description}

\begin{description}
\item[Telephoning Model] A fraction $loc\in[0,1]$ of the population is randomly selected at each step. Each individual in this subset interacts with a local neighbor. The remaining fraction of the population, $tel = 1 - loc$, chooses interaction partners globally.   
\end{description}
Note that $loc+tel=1$. This ensures that each agent has an opportunity to update their opinion once in a given time step. There is no relocation in the Telephoning Model. Note also that simultaneous updating means that attitudes are updated based on the spatial attitude configuration from the previous generation. The interactions with neighbors need not be reciprocal; even if the individual at $x$ chooses $z$, $z$ gets to choose its own interaction partner when deciding how to update.

We consider telephoning and relocation exclusively; that is, if $tel > 0$ then $rel = 0$, and if $rel > 0$ then $tel=0$. Consequently, every agent has one interaction per time step.  Relocation, which involves direct break-up of spatial structure by moving some individuals, is a standard mathematical way of introducing `mixing.'  Telephoning maintains the spatial locations of individuals over time, but allows some to interact with individuals outside the local neighborhood, thus breaking up some of the effects of spatial structure. In order to provide a fair comparison of Relocation and Telephoning, we do not use the influence functions in selecting global partners in the case of Telephoning. This mimics the ``influence-free'' choice of switching partners in the Relocation case. The influence functions apply only to local interactions.

The grid of attitudes is updated as follows. We update attitudes simultaneously: each agent picks another agent with whom to interact, determines how the focal agent's attitude should be updated according to the opinion of the other agent, and then implements the change at the next time step (once all other agents have determined how they should update their attitudes). At each time step we measure the distribution of attitudes.  We say a population is \textit{polarized} when the majority of the population is roughly balanced on the extreme ends of the attitude spectrum, a population is \textit{centered} when most attitudes reside in the center of the spectrum (e.g., on $-1$ and $1$), and a population reaches \textit{consensus} when everyone has the same opinion (i.e., all attitudes on the same side of the spectrum).


\section{The Well-Mixed (ODE) Case}
\label{wellmixedcase}

In \cite{baumgaertner2016opinion} we analyzed the effects of amplification with only local interactions, i.e., the fully spatial model with both amplification and spatial structure.  We found that amplification in combination with spatial structure promoted the clustering of like opinions and polarization towards more extreme attitudes. In this paper we are interested in the effects of amplification when interactions are global as well as local.  In the case where interactions are entirely global, the population is ``well-mixed" and spatial structure is removed. We begin by studying the well-mixed case where we can observe the effect of amplification in isolation from spatial structure.  In terms of the ABM, this case is achieved by setting telephoning or relocation to the maximum possible fraction ($tel=1$ or $rel=1$). 
In our Relocation and Telephoning models, this means that there are no local interactions occurring, and thus no influence (see above).

\subsection{ODE Approximation}

In the well-mixed case, each individual can interact with any individual in the entire population with equal probability.  An interaction between individuals occurs with a rate that depends on the frequencies of their attitudes in the population.  It is thus the frequency of each attitude that we track, and the evolution of these frequencies can be approximated using a system of ordinary differential equations.

\begin{figure}
\centering
\includegraphics[width=0.8\textwidth]{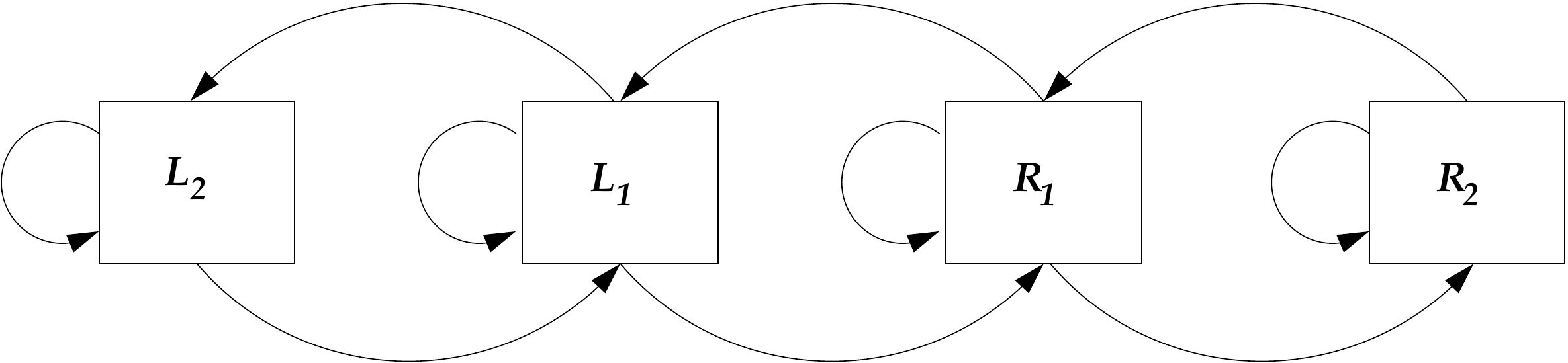}
\caption{Transition diagram for the case $L=2$.  The frequencies of the $a=-2$ and $a=-1$ subpopulations are $L_2$ and $L_1$, and the frequencies of the $a=1$ and $a=2$ subpopulations are $R_1$ and $R_2$.  The arrows show how each frequency is increased or decreased by individuals from one compartment of the population transitioning to another compartment.  So, for example, the $L_2$ subpopulation can only be increased by $L_1$ individuals becoming more entrenched, and not by $R_1$ or $R_2$ individuals transitioning directly to the $L_2$ subpopulation.  Note that interactions are not shown.}
\label{fig:ode-transitions}
\end{figure}

For simplicity, we consider here the case $L=2$. If we let the frequencies of attitudes $a=-2$ and $a=-1$ be represented as $L_{2}$ and $L_{1}$, and the frequencies of attitudes $a=1$ and $a=2$ be represented as $R_1$ and $R_2$, respectively, then we arrive at the transition diagram shown in Figure~\ref{fig:ode-transitions}.  Note that no attitude level can be ``skipped'' as attitudes change.  That is, individuals with attitude $a$ can only switch to neighbouring attitudes $a-1$ and $a+1$, or remain at $a$.
The rates for each transition in Figure~\ref{fig:ode-transitions} depend on the interactions, which are not shown.  Recall that in Figure~\ref{fig:amp_examples}, in order to determine the outcome of the interaction, it was necessary to identify a focal individual (the individual whose attitude would be changed by the interaction) and an interaction partner (the individual whose attitude would not be changed by the interaction).  We can think of Figure~\ref{fig:ode-transitions} as only showing the focal individuals.  The transition rates (coefficients for each arrow) depend on the interaction partners.  Thus, the $L_1$ to $L_2$ transition occurs when $L_1$ focal individuals interact with $L_2$ individuals, or with $L_1$ individuals in the presence of amplification.  All other interaction partners ($L_1$ without amplification, $R_1$, and $R_2$) will result in either an $L_1$ to $R_1$ transition, or no transition at all.  By tracking how all of the possible interactions contribute to the transitions in Figure~\ref{fig:ode-transitions}, we arrive at the following system of differential equations:  
\begin{subequations}
\begin{alignat}{3}
&\dot L_{2} & \quad & =  & \quad & L_{1}[L_{2}+ p_a L_{1}] - L_{2}[1- L_{2} -p_a L_{1}], \label{ODE-1} \\ 
&\dot L_{1} && = && L_{2}[1- L_{2} -p_a L_{1}] 
+ R_{1}[L_{2}+L_{1}] 
- L_{1}[1-(1-p_a)L_{1} ], \label{ODE-2}\\
&\dot R_{1} && = && L_{1}[R_{1}+R_{2}] + R_{2}[1-R_{2} -p_a R_{1}] - R_{1}[1-(1-p_a) R_{1} ], \label{ODE-3} \\
&\dot R_{2} && = && R_{1}[R_{2}+ p_a R_{1}] - R_{2}[1- R_{2} -p_a R_{1}] \label{ODE-4} .
\end{alignat}
\label{eq:ODEsimplified}
\end{subequations}
The full derivation can be found in Appendix~\ref{app:ODEderivation}.


\subsection{Time to Consensus under High Mixing}

\begin{figure}
\centering
\includegraphics[width=0.8\textwidth]{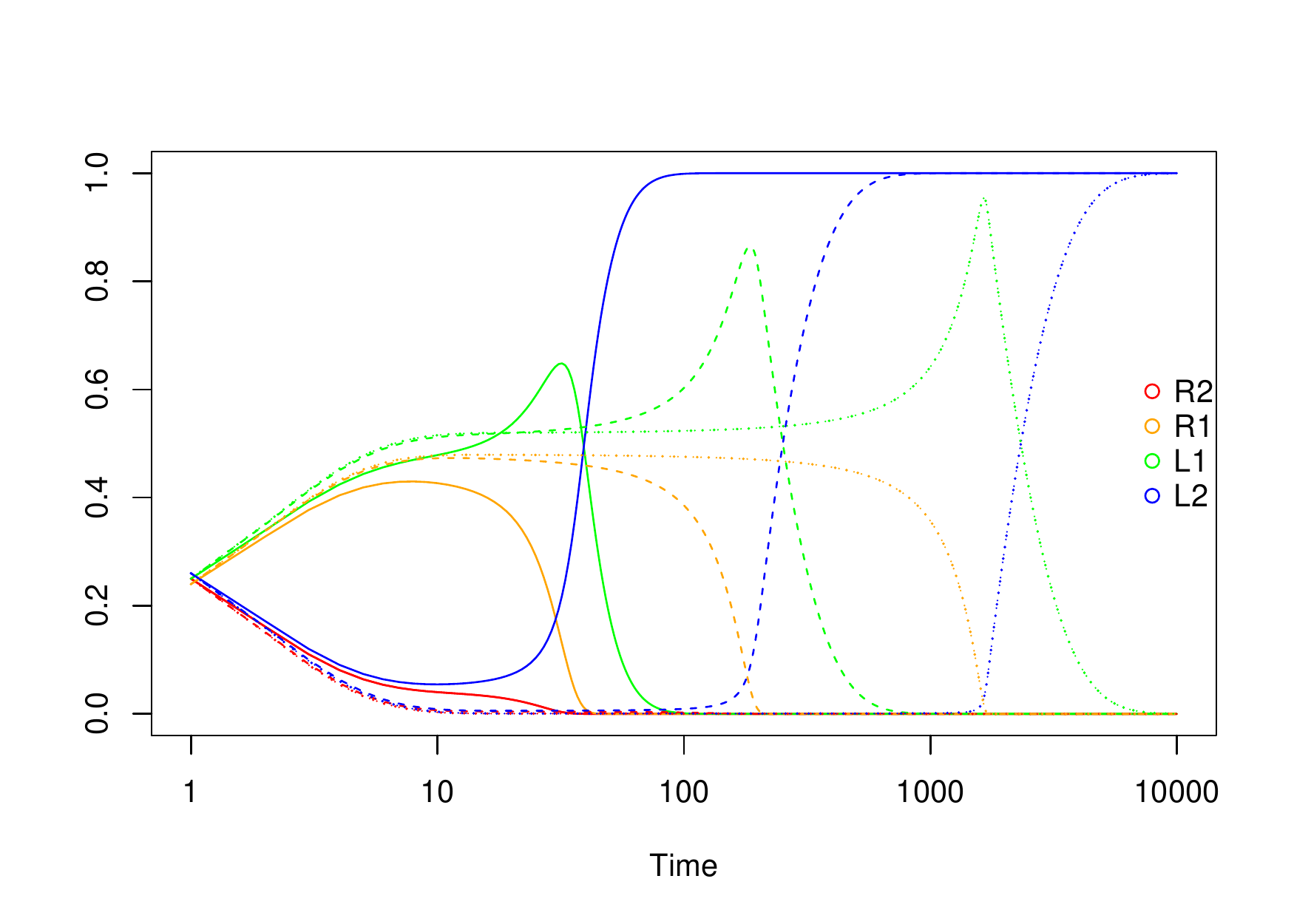}
\caption{(Color online) Solution to the ODE model~\eqref{eq:ODEsimplified} for three levels of amplification: $p_a=0.1$ (solid),  $p_a=0.01$ (dashed), and  $p_a=0.001$ (dotted).  ``$R$" refers to right (positive) opinions, ``$L$" refers to left (negative) opinions, and numbers indicate level of entrenchment (thus ``$L_2$'' corresponds to attitude $a=-2$, etc.).  Initial conditions were $R_2=0.25$, $R_1=0.24$, $L_1=0.25$, $L_2=0.26$.  The time axis is in units of $\log$(time). Note that consensus is reached as soon as the frequency of one opinion type reaches zero, in this case, when $R_1+R_2=0$. Dynamics after consensus are shown for visual purposes.}
\label{fig:ODEsims_prelim_uniformI}
\end{figure}

Numerical solutions of~\eqref{eq:ODEsimplified} show how time to consensus varies with amplification.  Specifically, 
Figure \ref{fig:ODEsims_prelim_uniformI} shows how a decrease in amplification by an order of magnitude (e.g., from 0.1 to 0.01) causes the time to consensus to increase by roughly an order of magnitude in the ODE.  Moreover, the maximum frequency of an inner opinion also increases as the amplification is decreased.

\begin{figure}
\centering
	\begin{subfigure}[t]{0.34\textwidth}
	\includegraphics[width=\textwidth]{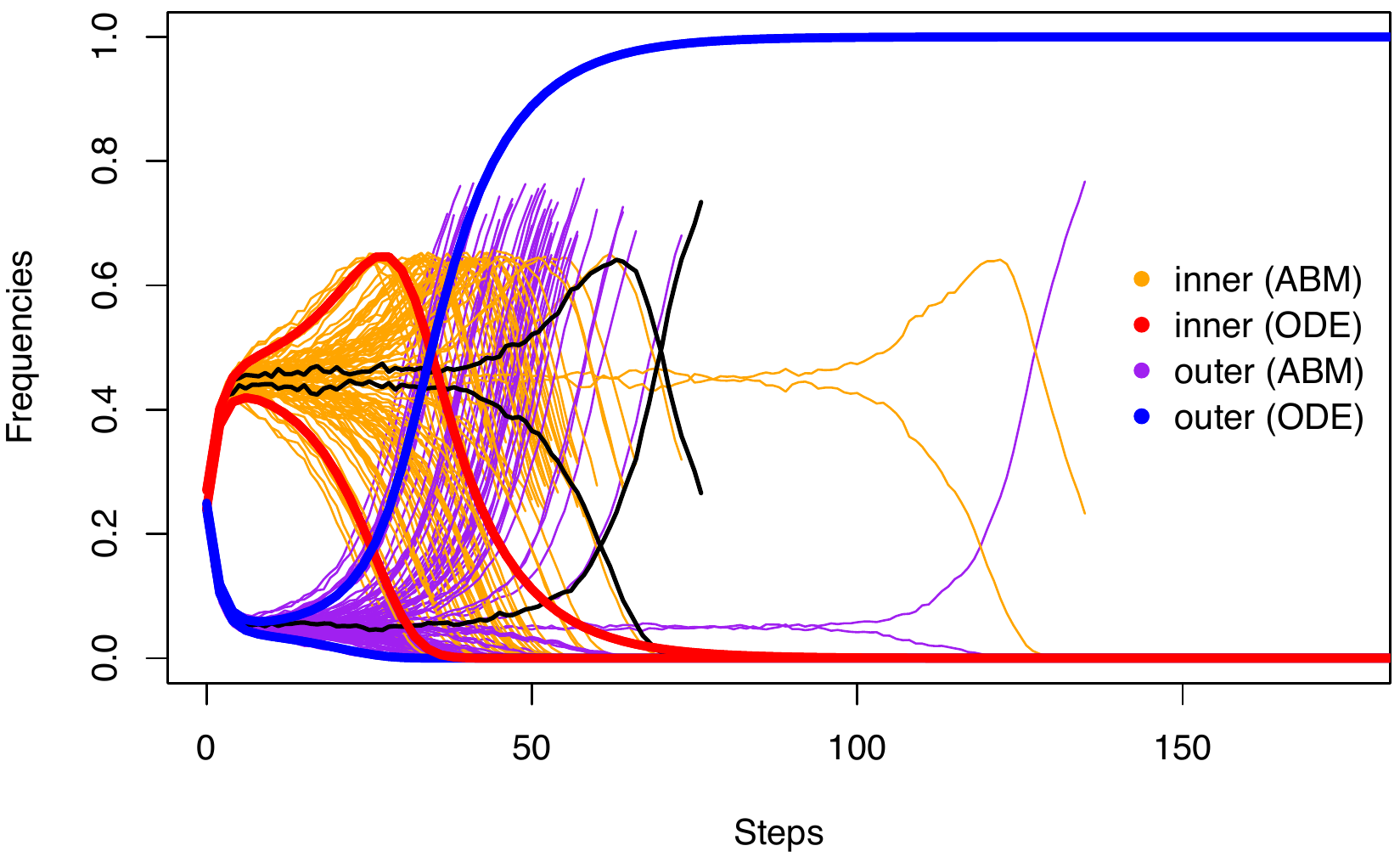}
	\caption{Amp = 0.1}
    \end{subfigure}
    \begin{subfigure}[t]{0.32\textwidth}
    \includegraphics[width=\textwidth]{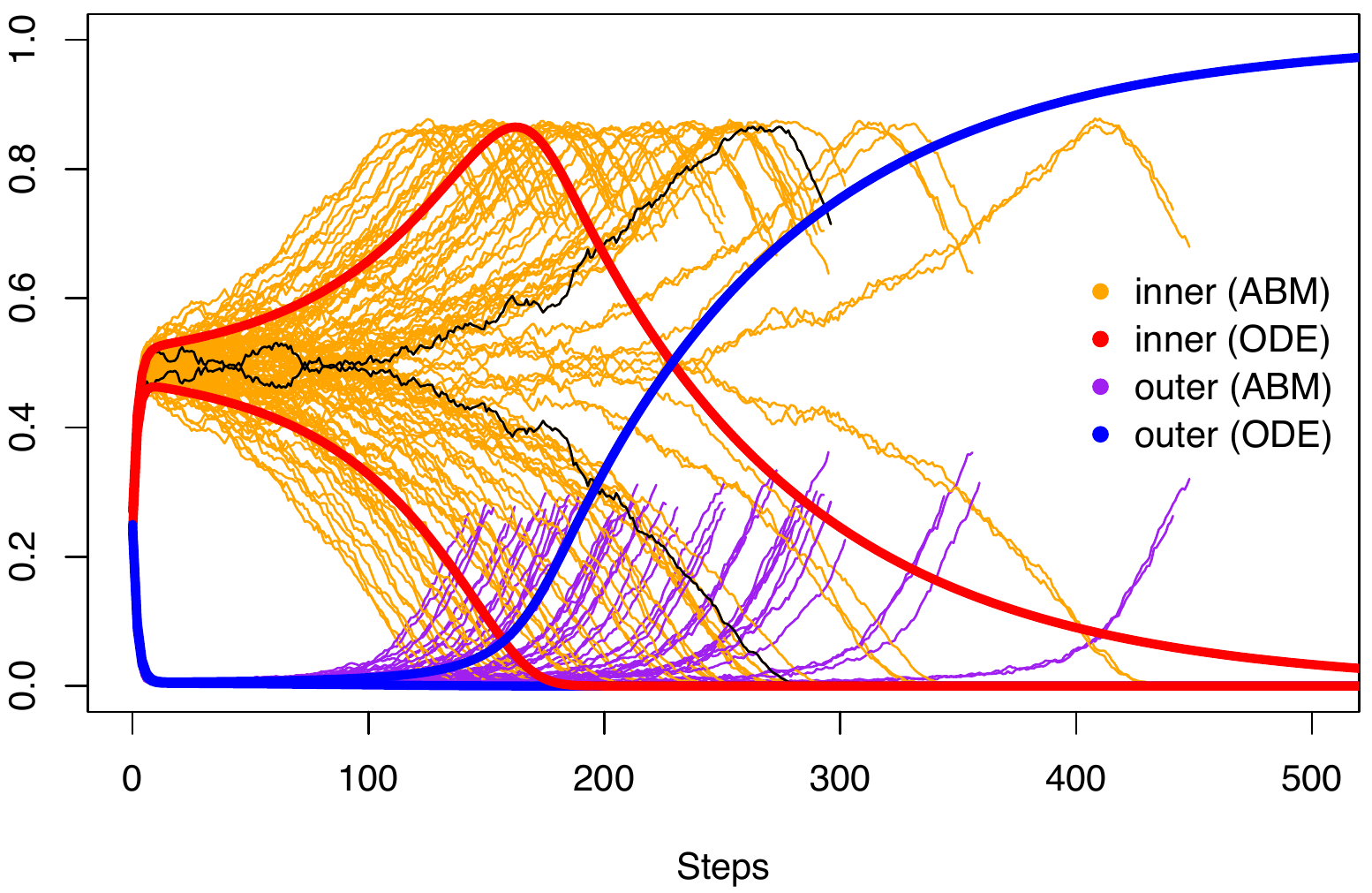}
    \caption{Amp = 0.01}
    \end{subfigure}
    \begin{subfigure}[t]{0.32\textwidth}
    \includegraphics[width=\textwidth]{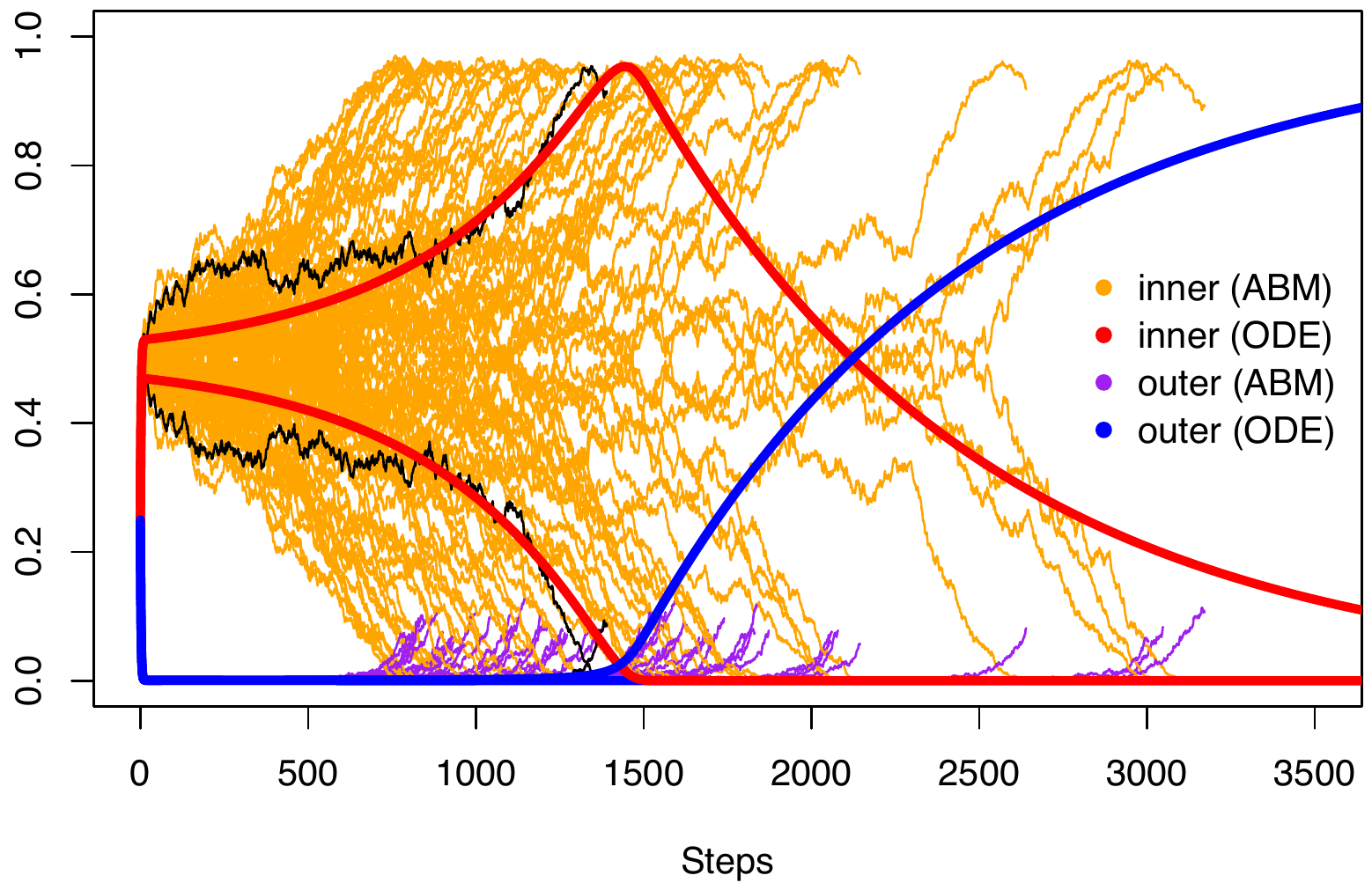}
    \caption{Amp = 0.001}
    \end{subfigure}
        \begin{subfigure}[t]{0.32\textwidth}
    \includegraphics[width=\textwidth]{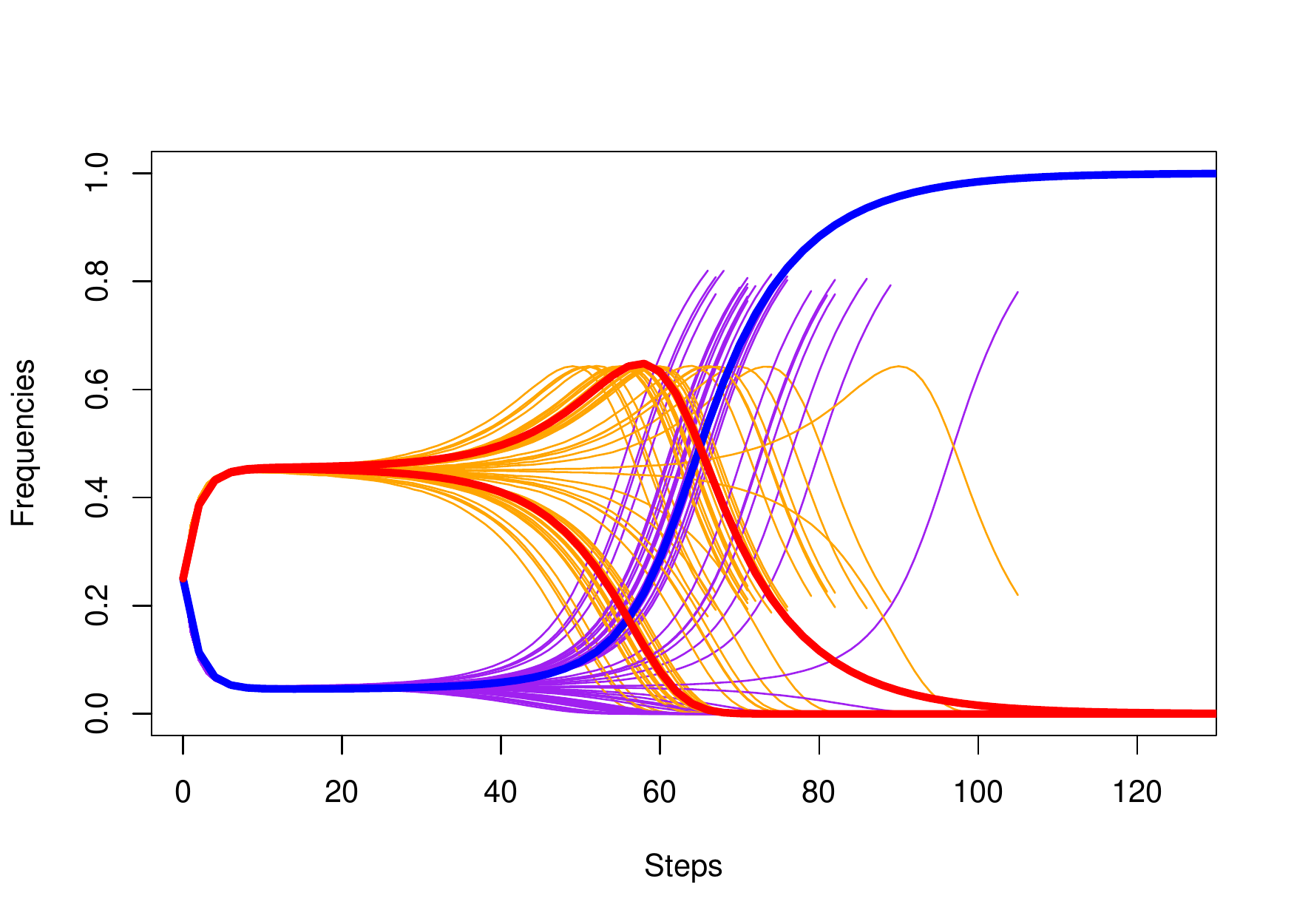}
    \caption{Amp = 0.1}
    \end{subfigure}
        \begin{subfigure}[t]{0.32\textwidth}
    \includegraphics[width=\textwidth]{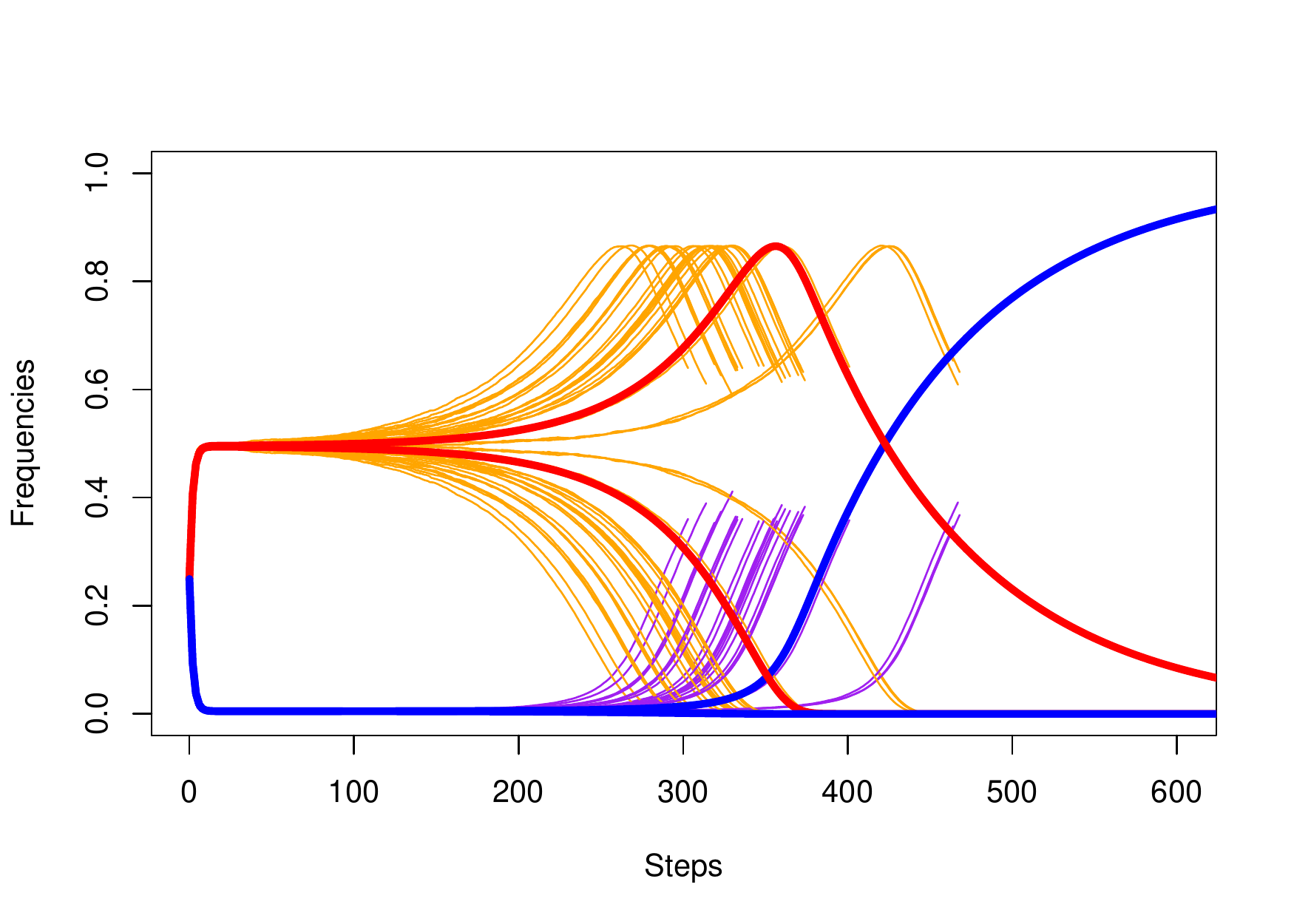}
    \caption{Amp = 0.01}
    \end{subfigure}
        \begin{subfigure}[t]{0.32\textwidth}
    \includegraphics[width=\textwidth]{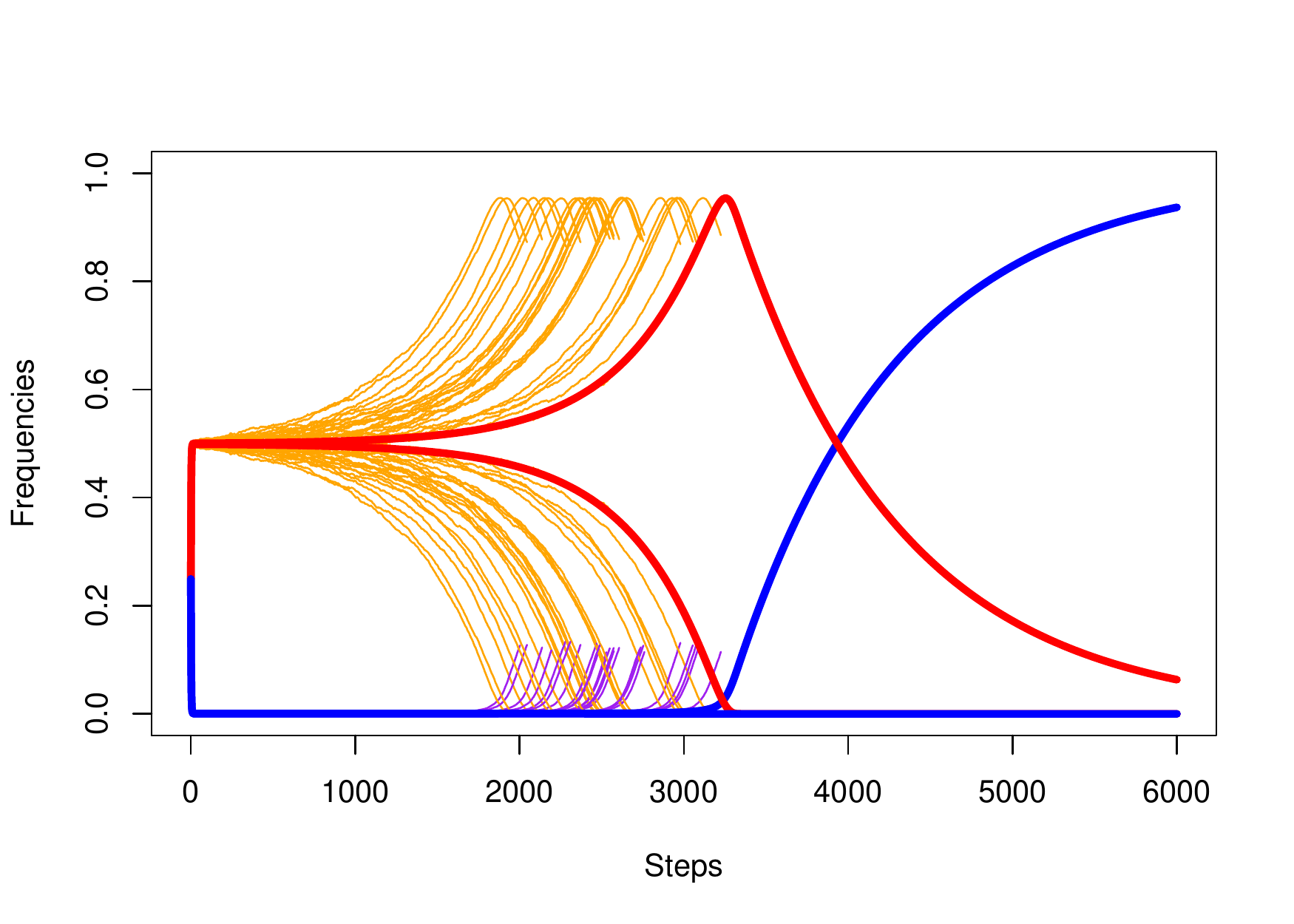}
    \caption{Amp = 0.001}
    \end{subfigure}
\caption{\label{fig:ode_abm} (Color online) Plots show changes in attitude frequencies as populations reach consensus. The top row is for population size $N=10,201$ and the bottom row is for $N=750,000$.  Consensus is reached when the inner and outer opinions of one type are zero, e.g., $L_1+L_2=0$. Thick, smooth lines show the numerical solutions of the ODE. Thin, stochastic lines show the simulation results with Telephoning at 100\%. Black lines highlight one specific simulation. Three levels of amplification were used, 0.1, 0.01, and 0.001. In each scenario the simulations closely matched the ODE predictions in two ways: one is that the time to consensus grows as amplification decreases (note the change in timescale from (a) to (b) to (c), and also from (d) to (e) to (f)), the second is the maximum height of the inner opinions. As the differences in the rows indicate, smaller population sizes contribute to stochasticity, larger less so. Because the ABM version is stochastic, we used the average of the initial conditions of the ABM simulations for the initial conditions of the ODE.}
\end{figure}

Simulations of the (agent-based) Relocation and Telephoning models with maximum mixing are consistent with these predictions. Figure \ref{fig:ode_abm} shows the simulations of the Telephoning model for three levels of amplification and compares them with the ODE numerical solutions. The deterministic ODE system provides a good approximation of the stochastic agent-based system for predicting the qualitative shape of the frequency curves, including the peak frequency of the inner opinion before consensus.

The solutions in Figure~\ref{fig:ode_abm} also reveal characteristic behaviors of the well-mixed system.  In all of the simulations shown, attitudes are initially distributed uniformly throughout the population.  When interactions begin, there is initially a very rapid increase in the moderate (inner) attitudes, $-1$ and $1$, and a parallel decrease in the extreme (outer) attitudes, $-2$ and $2$.  This rapid centering is then followed by a quasi-equilibrium where the centered distribution changes only slowly.  Finally, one of the extreme attitudes begins to increase rapidly (in Figure~\ref{fig:ode_abm} it is the $+2$ attitude that increases) while the other attitudes all decrease.  The ODE model can be used to explain this sequence of behaviors.


\subsection{ODE Model Analysis} 
\label{sec:analysis}

The solution behavior shown in Figure~\ref{fig:ODEsims_prelim_uniformI} has four distinct parts: (1) the initial rapid centering, (2) the period of pseudo-stability at the centered state, (3) the eventual symmetry-breaking that leads to (4) consensus on one extreme opinion.  To show how these behaviors arise, we study the steady states of the ODE model~\eqref{eq:ODEsimplified}, and then study the phase plane of two sub-models derived through simplifications of the original model.

The steady states of the model~\eqref{eq:ODEsimplified} that satisfy the constraint $L_{2}+L_{1}+R_1+R_2=1$, are
\begin{alignat*}{2}
& \text{(i) } (0,\alpha,1-\alpha,0), & & \\
& \text{(ii) } (0,0,1-\alpha,\alpha), & \quad & \text{(iii) } (\alpha,1-\alpha,0,0), 
\end{alignat*}
where the arbitrary value $\alpha\in[0,1]$.  Steady state (i) corresponds to a completely symmetrical centered state when $\alpha=1/2$. Steady states (ii) and (iii) correspond to consensus.  Consensus on an extreme opinion occurs when $\alpha=1$.  All of these steady states are saddle nodes. In order to lend insight into the system behavior, we make some simplifying assumptions to derive two sub-models that are amenable to steady-state analysis. The first sub-model is relevant during centering (Section~\ref{ssec:analysis-centering}), during the pseudo-stable behavior at the centered state (Section~\ref{ssec:analysis-centre-pseudo-steady}), and during symmetry-breaking (Section~\ref{ssec:analysis-symmetry-breaking}).  The second sub-model applies to the transition from symmetry-breaking to consensus (Section~\ref{ssec:analysis-consensus}).

\subsubsection{Centering}
\label{ssec:analysis-centering}

\begin{figure}
\centering
\includegraphics[width=0.7\textwidth]{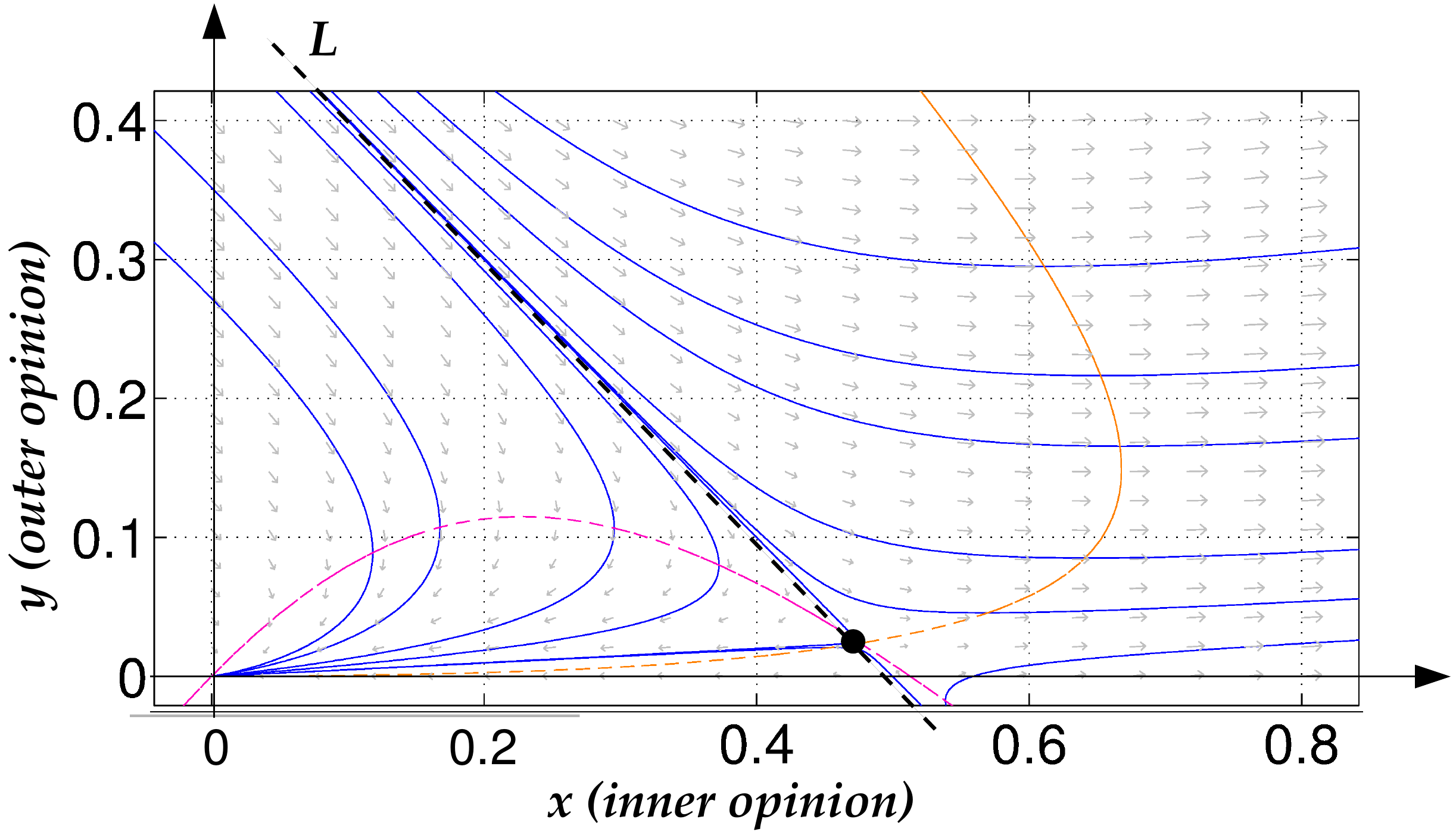}
\caption{(Color online) Phase plane plot for equations~\eqref{eq:ODE_2D_centering} with $p_a=0.05$.  The nullclines (curves along which $dx/dt=0$ or $dy/dt=0$) are shown in orange and magenta.  The coexistence steady state is indicated by a black dot.  Solutions satisfying the symmetry condition $L_2=R_2=y$, and $L_1=R_1=x$ are constrained to the line $x(t)+y(t)=1/2$, which appears as a dashed black line (labeled $S$). Solution trajectories $(x(t),y(t))$ of~\eqref{eq:ODE_2D_centering} are shown in blue.}
\label{fig:ODE_2D_centering}
\end{figure}

The centering behavior that we identified in Figure~\ref{fig:ode_abm} arises within highly symmetric solutions where $L_{2}=R_2$ and $L_{1}=R_1$ (see Figure~\ref{fig:ODEsims_prelim_uniformI}).  If we let the outer opinions satisfy $L_2=R_2=y$, and the inner opinions satisfy $L_1=R_1=x$, equations~\eqref{eq:ODEsimplified} reduce to a two-dimensional system in $x$ and $y$:
\beqsub
\beqa
\frac{dx}{dt} & = & -y^2 - ((p_a-1)x - 1)y + ((2-p_a)x-1)x, \\
\frac{dy}{dt} & = & y^2 + ((p_a+1)x - 1)y + p_a x^2.
\eeqa
\label{eq:ODE_2D_centering}
\eeqsub
Figure~\ref{fig:ODE_2D_centering} shows the phase plane for equations~\eqref{eq:ODE_2D_centering} in the case where $p_a$ is small (the larger $p_a$ case is discussed in Section~\ref{ssec:large-pa}).
We observe that the system has two steady states, one at (0,0), and a coexistence state close to (1/2,0).  We can show (see \ref{app:stable-manifold}) that the stable manifold of the coexistence state is, in the case of $p_a=0$, exactly equal to the line $x+y=1/2$.  For $p_a>0$, the stable manifold is no longer exactly the line $x+y=1/2$, but it is very close as long as $p_a$ remains small.  

Trajectories $(x(t),y(t))$ show how the frequency of inner and outer opinions evolve in this phase plane.  The relevant trajectories are those that start on the line $x(0)+y(0)=1/2$ and, in particular, the one where $x(0)=y(0)=1/4$.  Since this line is close to the stable manifold of the steady state, the solution trajectory remains near the stable manifold, and approaches the steady state.  On this trajectory, the frequency of the outer(inner) opinion decreases(increases), and we observe centering.  In sum, for any initial condition with $x+y=1/2$, the solution trajectory will initially move toward the coexistence steady state near (1/2,0).  This explains the centering part of the solutions.

\subsubsection{The Center as a Pseudo Steady-State}
\label{ssec:analysis-centre-pseudo-steady}

The coexistence steady state of the symmetric system~\eqref{eq:ODE_2D_centering} is a saddle, which means that it is ultimately unstable.  Any solution trajectory that starts exactly on the stable manifold will terminate at the coexistence steady state and, in the absence of noise, remain there for all time.  Numerical solutions however, always contain small errors, and so eventually there will be a sufficient accumulation of these errors to cause the solution to veer away from the steady state.  The amount of time spent at the steady state depends on the distance between the initial conditions and the stable manifold: The smaller this distance, the longer the time spent at the coexistence steady state. Simulations verify this result (not shown).

The approach to the centered state and time spent there also depends on $p_a$.  As $p_a$ increases, the stable manifold moves further away from the line $x+y=1/2$, and so solution trajectories starting on that line do not end up as close to the centered steady state (phase plane not shown).

This analysis explains the pseudo-stability at the centered state.

\subsubsection{Symmetry-Breaking}
\label{ssec:analysis-symmetry-breaking}

Once the solution trajectory veers away from the steady state, the off-manifold trajectories (blue lines) become relevant.  If a perturbation of the trajectory takes it to a point below the stable manifold, the frequencies of both the inner and outer opinions rapidly approach zero.  If, on the other hand, a perturbation of the trajectory takes it to a point above the stable manifold, the frequency of the outer opinion remains small while that of the inner opinion rapidly increases.  As the trajectories move away from the stable manifold, however, equations~\eqref{eq:ODE_2D_centering} cease to be relevant.  Recall that the frequencies of all four opinions must add to 1.  In order for this constraint to be satisfied after the solution has been perturbed away from the line $x(t)+y(t)=1/2$, the frequencies of the two inner and two outer opinions can no longer be the same.  More specifically, if the system is perturbed to a point below the stable manifold, the left pair (say) of inner and outer opinion frequencies (i.e. $L_{1}$ and $L_{2}$) is rapidly approaching zero, which means that the sum of the right pair of inner and outer opinion frequencies (i.e. $R_1$ and $R_2$) must be approaching 1.  This situation violates the symmetry assumption ($L_{1}=R_1=x$, and $L_2=R_2=y$).  We thus have the mechanism for symmetry-breaking in the solution.  

Following the trajectories above the stable manifold, we see that changes in $y(t)$ are very small compared with changes in $x(t)$ (the trajectories move in a mostly horizontal direction away from the steady state), and so we expect that symmetry-breaking should be most evident in the inner opinion initially. This behavior can be observed in simulations (see, e.g., Figures~\ref{fig:ODEsims_prelim_uniformI} and \ref{fig:ode_abm}).

\subsection{Consensus}
\label{ssec:analysis-consensus}

\begin{figure}
\centering
\includegraphics[width=0.7\textwidth]{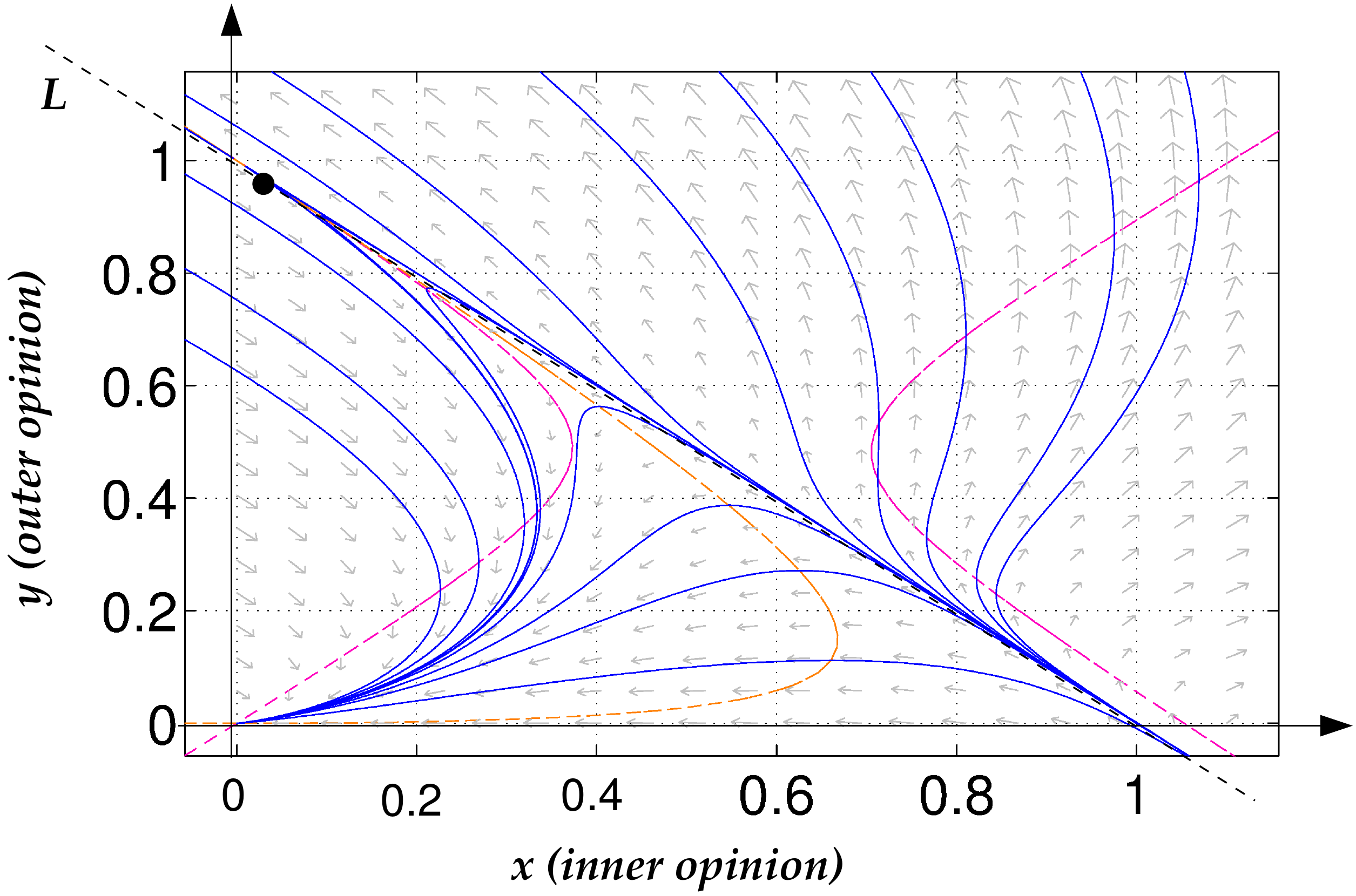}
\caption{Phase plane plot for~\eqref{eq:ODE_2D_consensus} with $p_a=0.05$.  The nullclines are shown in orange and magenta.  The steady state is in the vicinity of the black dot. Note that for $p_a>0$, we have $x>0$ at the steady state, indicating that the frequency of inner opinions does not become zero.  The dashed black line labeled $Q$ is the line $x_r+y_r=1$.  Strictly speaking, the dynamics of the full opinion model should occur only along the line $Q$ (and only for $0<x_r,y_r<1$), once the frequencies of the two opinions on the other side of the spectrum have been reduced to zero.  The full phase plane lends insight into the stability of the solution behavior.}
\label{fig:ODE_2D_consensus}
\end{figure}

Once symmetry-breaking has occurred, the left opinion frequencies (say) move rapidly toward zero, while the sum of the right opinion frequencies move rapidly away from zero.  We can thus write a new simplification of the model~\eqref{eq:ODEsimplified} in which the left opinion frequencies are zero, i.e., one opinion is lost. Let $L_1=L_2=0$, and name the remaining inner and outer opinions as $R_1=x_r$ and $R_2=y_r$.  Substituting these variables into~\eqref{eq:ODEsimplified} we arrive at the second sub-model 
\beqsub
\beqa
\frac{dx_r}{dt} & = & y_r(1-y_r-p_a x_r) - x_r(1-(1-p_a) x_r), 
\label{eq:ODE_2D_consensus_x}
\\
\frac{dy_r}{dt} & = & x_r(y_r+p_a x_r) - y_r(1-y_r-p_a x_r),
\label{eq:ODE_2D_consensus_y}
\eeqa
\label{eq:ODE_2D_consensus}
\eeqsub
\noindent The model applies equally to the situation where the roles of left and right are reversed (i.e. setting $R_1=R_2=0$, $L_1=x_l$, and $L_2=y_l$, we arrive at~\eqref{eq:ODE_2D_consensus} with $x_r$ and $y_r$ replaced with $x_l$ and $y_l$ respectively).
The phase plane diagram for equations~\eqref{eq:ODE_2D_consensus} with $p_a=0.05$ is shown in Figure~\ref{fig:ODE_2D_consensus}.  The dynamics being illustrated here are the ones that occur when both the inner and outer opinions on one side of the spectrum have dropped to zero, and so the frequency of the remaining two opinions should add up to 1.  

Solutions of equations~\ref{eq:ODE_2D_consensus} should move along the line $x_r+y_r=1$.  We observe that the solution direction along $x_r+y_r=1$ in Figure~\ref{fig:ODE_2D_consensus} is from the lower right, where the frequency of the inner opinion is near 1 but the frequency of the outer opinion is near 0, to the top left, where the values of the two frequencies are reversed.  Thus, the solutions of~\eqref{eq:ODEsimplified} eventually move toward the $(0,0,\alpha,1-\alpha)$ (or $(1-\alpha,\alpha,0,0)$) steady state.  The size of $\alpha$ decreases toward 0 as $p_a$ also decreases toward 0. The final state is thus consensus on the right (or left), with the frequency of outer opinions dominating the solution at a value close to 1.

\subsection{The Large Amplification Case}
\label{ssec:large-pa}

When the amplification probability $p_a$ is not small the previous analysis still applies, but the duration of the pseudo-stable behavior at the centered state decreases as $p_a$ increases.  Eventually it becomes difficult to distinguish transition points between the three initial behaviors.

\section{Structured Populations with Some Mixing}
\label{results}

We have analyzed two ends of the mixing continuum: the case where there is no population structure (complete mixing; see Section \ref{wellmixedcase}) and the case where there is no mixing (details in \cite{baumgaertner2016opinion}). We will refer generically to the probabilities of relocation or telephoning as the ``level of mixing." Here we analyze and compare the relocation model and telephoning model, paying particular attention to cases where they differ. The model can be downloaded from the NetLogo modeling commons: \url{http://modelingcommons.org/browse/one_model/4963#model_tabs_browse_info} (It requires a particular version of the \texttt{rnd} extension, which can be found as an additional file in the modeling commons.) 

Simulating our agent-based model, we found that for moderate to higher levels of mixing (e.g., either $rel\geq 0.25$ or $tel\geq 0.25$) there is little difference between relocation and telephoning, and that the ODE system described in Section \ref{wellmixedcase} remains a good approximation of the dynamics of our agent-based system (results not shown). Deviations from the ODE approximation arise for lower levels of mixing (e.g., either $rel\leq 0.1$ or $tel\leq 0.1$). For such settings, we also see differences in the effects of relocation and telephoning.

\subsection{Consensus Times}

A common summary statistic for opinion dynamics models is time to consensus. In our model, consensus refers to everyone in the population having the same opinion, though it is possible they differ in attitude. 
That is, all attitudes have the same sign (positive or negative) but can differ in strength (i.e., magnitude). 

\begin{figure}
\centering
\includegraphics[width=\textwidth]{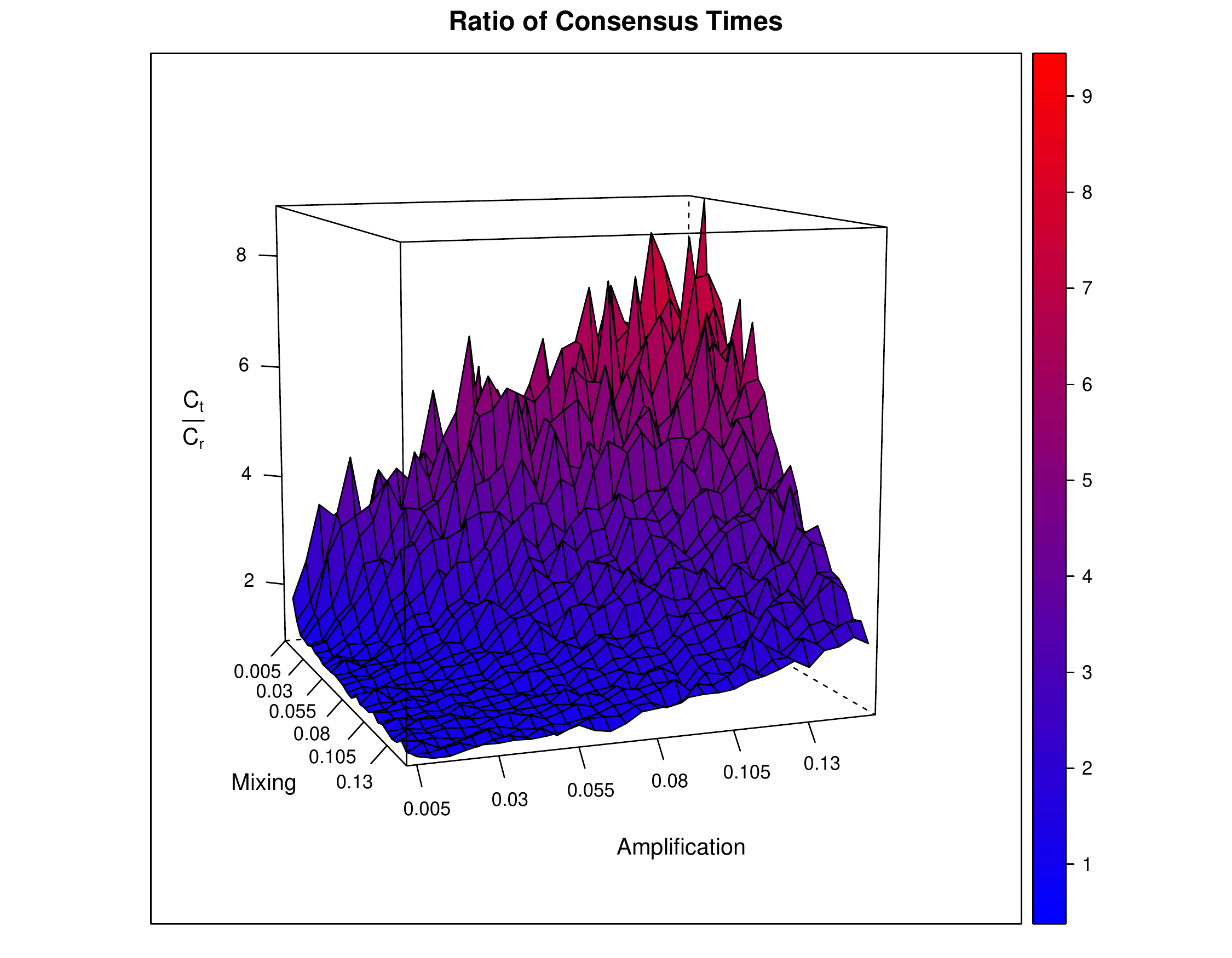}
\caption{\label{relativetime} (Color online) Comparison of mean consensus times under telephoning and relocation. The surface is based on 50 simulations for each point in the parameter space for each of telephoning and relocation. For each point, the plot shows $\frac{C_t}{C_r}$, where $C_t$ and $C_r$ are the average consensus times for telephoning and relocation, respectively. When amplification is high and mixing is low (the back corner), telephoning takes about five to eight times longer than relocation  to produce consensus. When amplification is low and mixing is high (the front corner) the times to consensus for telephoning and relocation are about the same. When amplification is low and mixing is low (the left corner), or when both amplification and mixing are high (right corner), telephoning takes about twice as long as relocation. Sample points on both axes range from 0.005 up to 0.15  in increments of 0.005.}
\end{figure}

We simulated our agent-based model under various degrees of amplification and mixing (see Figure \ref{relativetime}).  We found that relocation and telephoning differ significantly in their consensus times for low levels of mixing. In particular, when using comparable probabilities of relocation or telephoning (only one at a time), the consensus times for telephoning can take several times longer than those of relocation. This difference increases as amplification increases, even at lower levels of amplification. 

\begin{figure}
\centering
\includegraphics[width=0.9\textwidth]{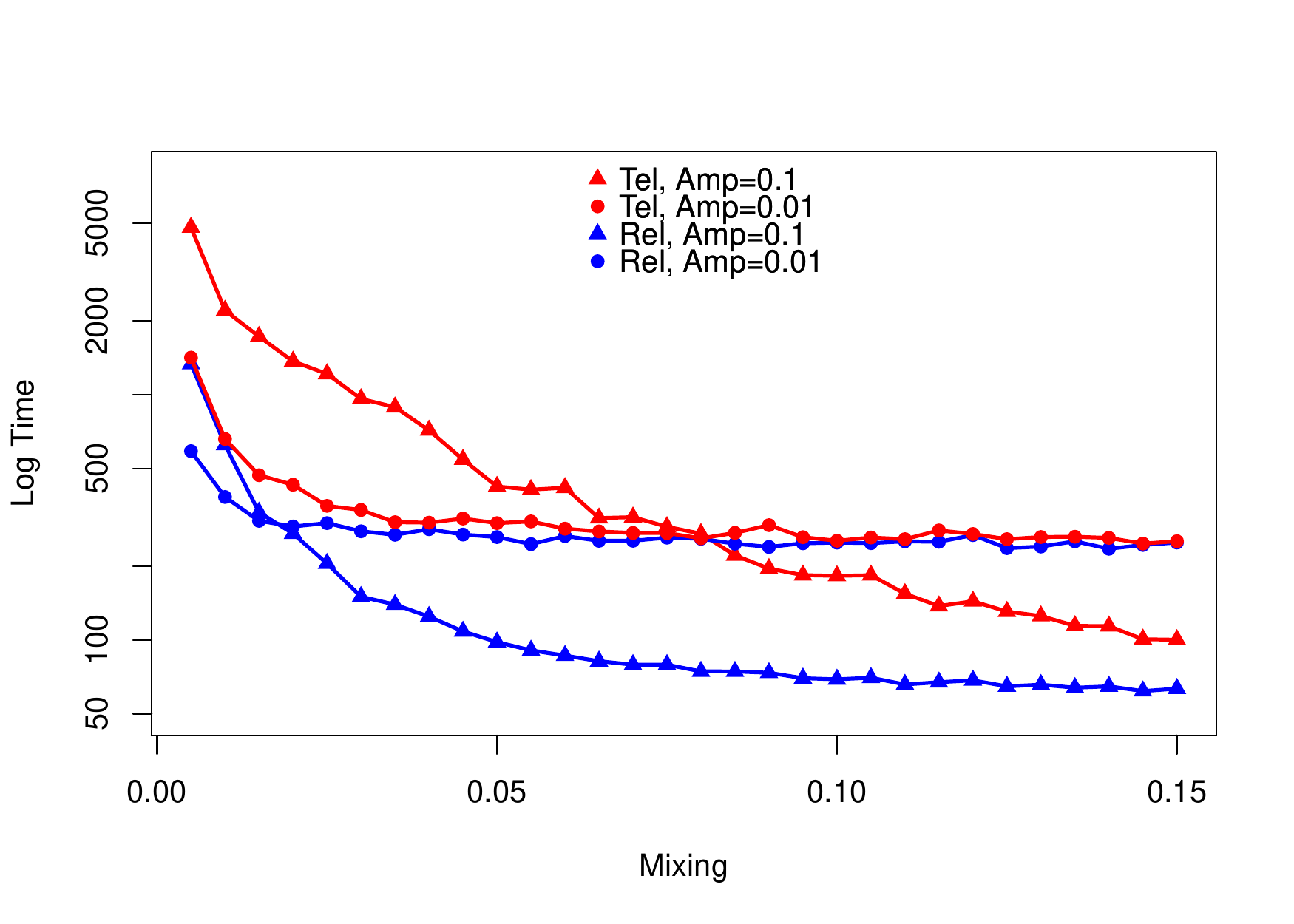}
\caption{\label{timelogplot} (Color online) Mean consensus times (on a log scale) as a function of mixing level for the case of uniform influence for local interactions (50 simulations for each point). Data are plotted for relocation and telephoning, each with two levels of amplification, very low ($p_a=0.01$) and low ($p_a=0.1$). Note that for very low levels of amplification relocation and telephoning converge on similar consensus times early on, approximately $rel=0.03$ or $tel=0.03$ (see lines with circles for data points). When amplification is higher but still low ($p_a=0.1$) consensus times for telephoning are not only higher, but also take longer to decrease (compare red triangles to blue triangles). The first value of mixing is 0.005. The case of no mixing is not shown.}
\end{figure}

Figure \ref{timelogplot} shows the consensus times for relocation and telephoning for levels of amplification that are low or very low. When amplification is very low (say, $p_a=0.01$), consensus time decreases quickly as mixing increases from 0.005 to 0.03, but then stabilizes around 300 thereafter. On the other hand, when amplification probability is only low (say, $p_a=0.1$), consensus time continues to decrease as mixing is increased. The difference between telephoning and relocation is also more stark as amplification is increased and mixing is decreased. Not only is the consensus time for telephoning longer than relocation, but the rate at which it decreases as a function of mixing probability is less than relocation.  

It is worth noting the effects of amplification differ in highly structured populations (where there is effectively no mixing) and populations where there is some mixing. When a population is highly structured, consensus times increase as amplification is increased for both telephoning and relocation.  As more relocation is added (e.g., $rel=0.02$), this relationship reversed; increasing amplification decreases consensus time (see where line with blue triangles and line with blue circles intersect in Figure \ref{timelogplot}). A similar effect is observed for telephoning, but not until telephoning is around $tel>=0.08$ (see where line with red triangles intersects line with red circles in Figure \ref{timelogplot}). The explanation for this difference is related to the effects that the two modes of mixing have on spatial structure. As our analysis of the well-mixed case suggests, increasing amplification decreases times to consensus (Figure \ref{fig:ODEsims_prelim_uniformI}).  Hence, if telephoning `mixes' the population less than relocation, we expect that it will take more telephoning to produce similar consensus times as relocation (given some level of amplification). 

\begin{figure}
\centering
\includegraphics[width=0.9\textwidth]{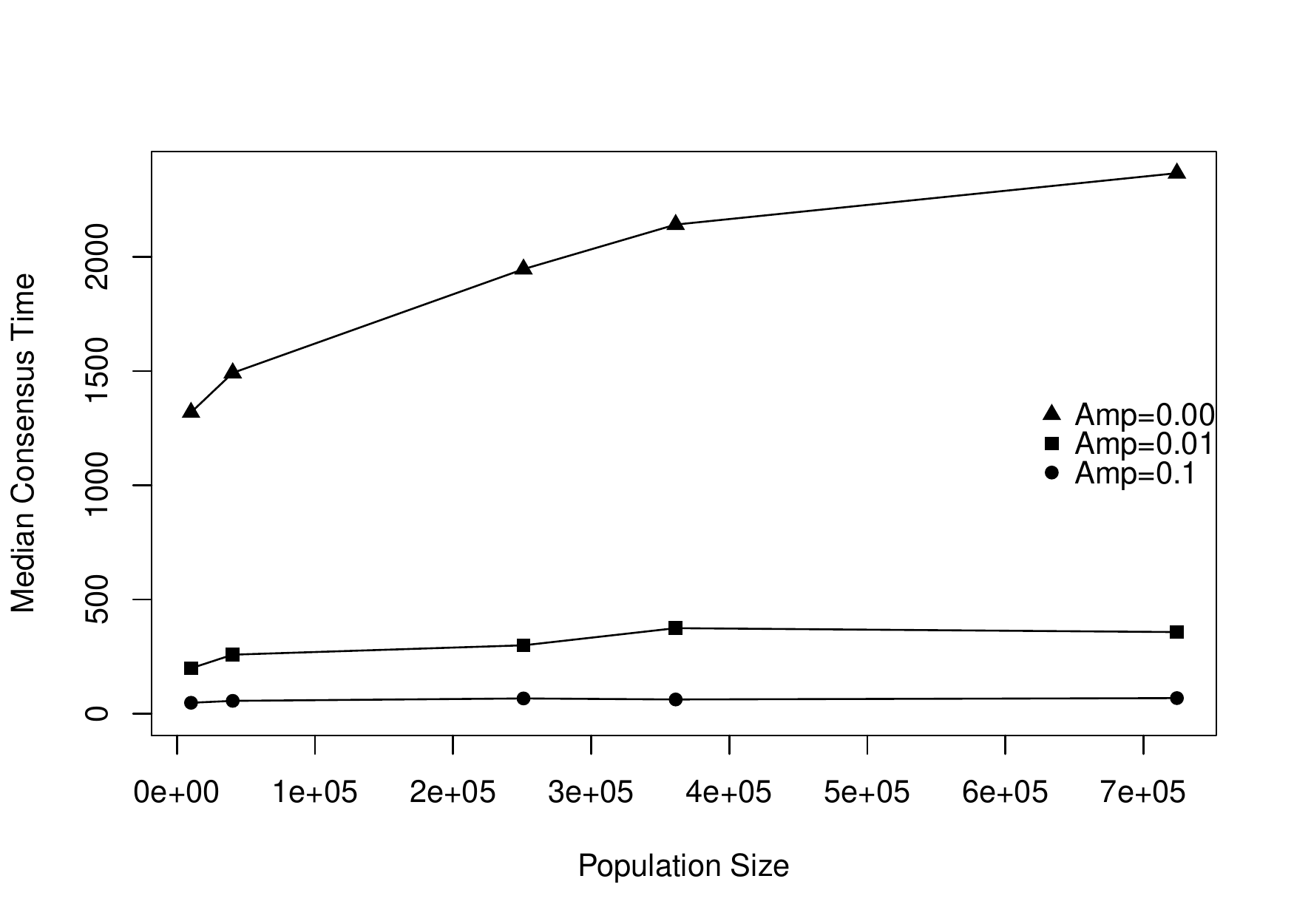}
\caption{\label{systemsize} Median consensus times relative to population size for different levels of amplification. Each point represents the median of 25 runs.}
\end{figure}

In addition, we tested the effect of system size on consensus times in the case of maximal mixing (Figure \ref{systemsize}). We find that system size has a relatively small effect and saturates as population size is increased. The predominant effect on consensus times is explained by the level of amplification, as indicated by the separation of the curves in Figure \ref{systemsize}. That is, as amplification is increased in our well-mixed scenarios, consensus times decrease (across population sizes). 

In the next section we analyze the spatial effects of relocation and telephoning in more detail.


\subsection{Spatial Behaviors}

In \cite{baumgaertner2016opinion} we found that, in the absence of mixing, our model can  produce clusters of opinions, and, when amplification is high enough, a certain amount of \textit{surface tension} and motion by mean curvature (i.e., opinions on the interior of a cluster would eventually get swallowed up by opinions on the exterior of a cluster). 
Here we examine the effects that relocation and telephoning have on clusters and their boundaries. 

\begin{figure}
\centering
\includegraphics[width=.15\textwidth]{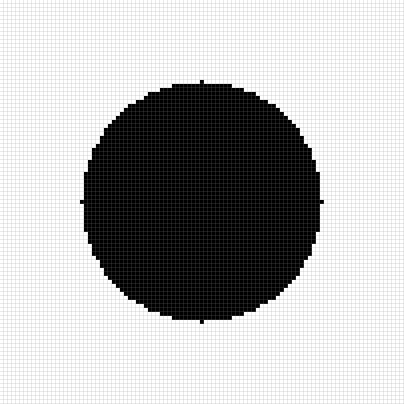}
\includegraphics[width=.15\textwidth]{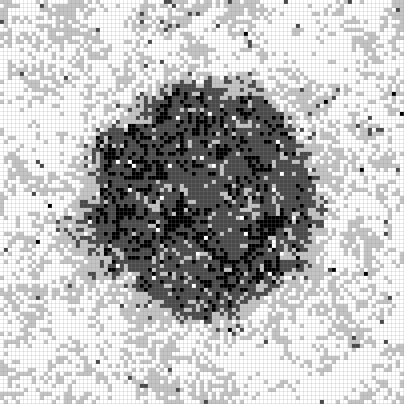}
\includegraphics[width=.15\textwidth]{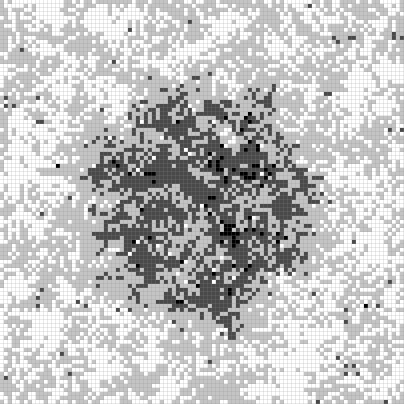}
\includegraphics[width=.15\textwidth]{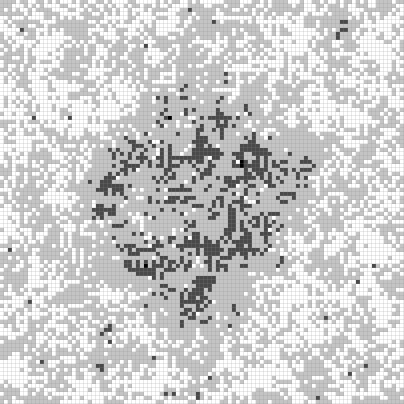}
\includegraphics[width=.15\textwidth]{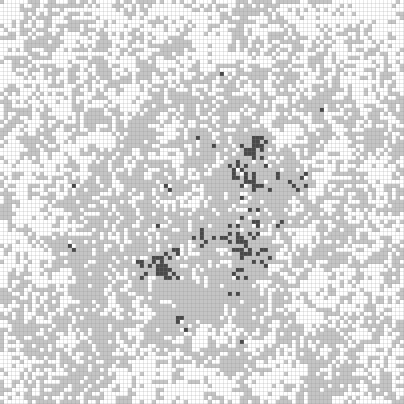}
\includegraphics[width=.15\textwidth]{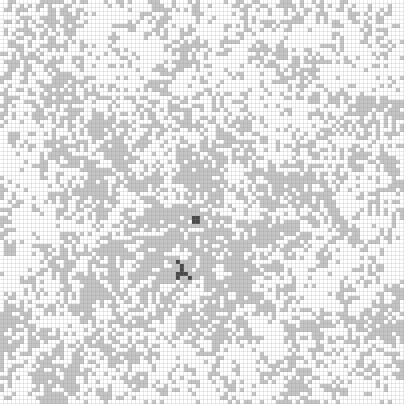}\\
\vspace{10pt}
\includegraphics[width=.15\textwidth]{CircleRelo0}
\includegraphics[width=.15\textwidth]{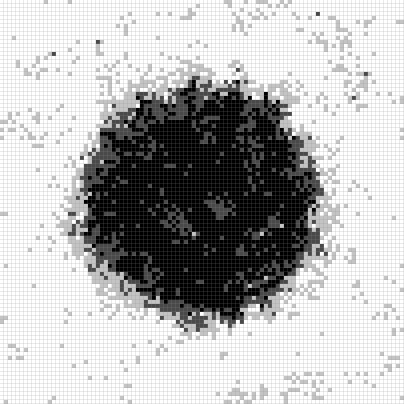}
\includegraphics[width=.15\textwidth]{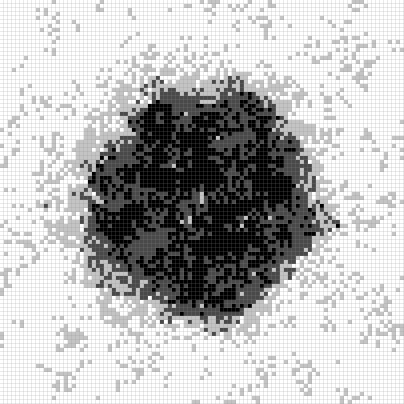}
\includegraphics[width=.15\textwidth]{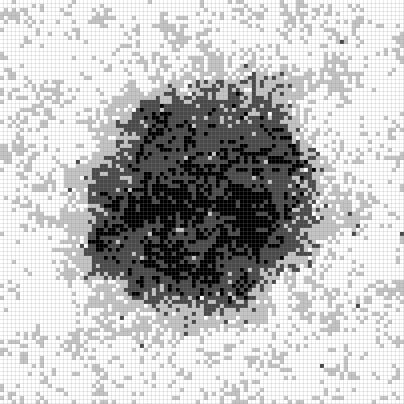}
\includegraphics[width=.15\textwidth]{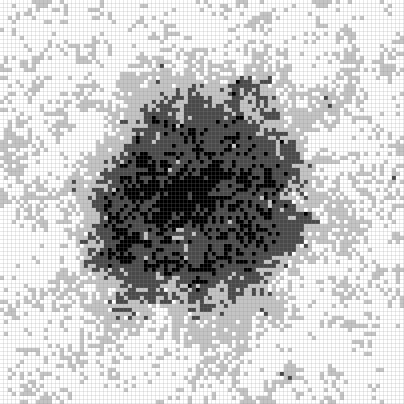}
\includegraphics[width=.15\textwidth]{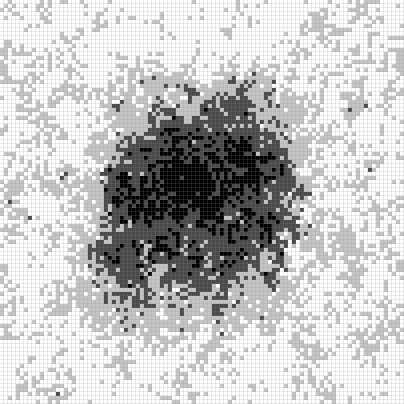}\\
\vspace{10pt}
\includegraphics[width=.15\textwidth]{CircleRelo0}
\includegraphics[width=.15\textwidth]{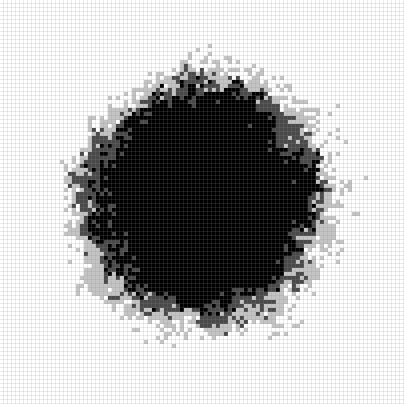}
\includegraphics[width=.15\textwidth]{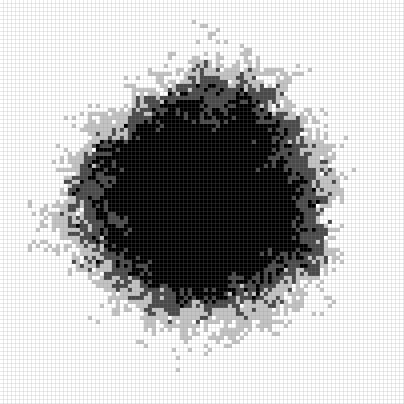}
\includegraphics[width=.15\textwidth]{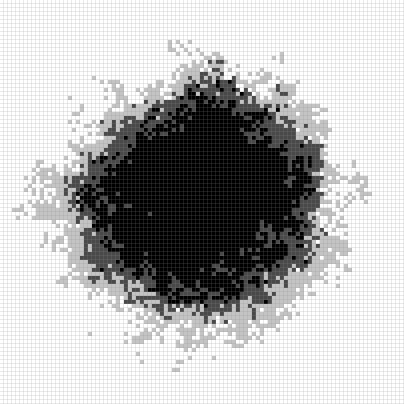}
\includegraphics[width=.15\textwidth]{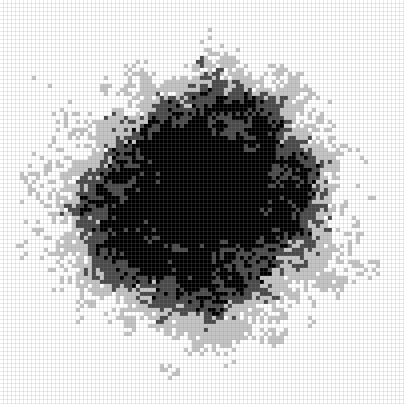}
\includegraphics[width=.15\textwidth]{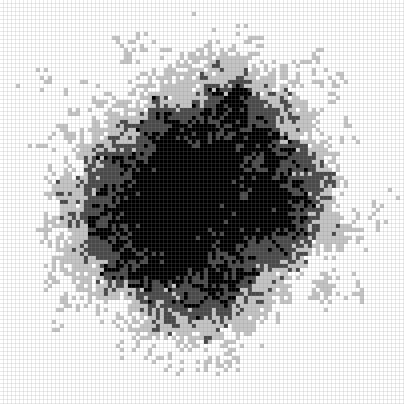}

\caption{\label{consensus} Evolution of attitudes over 125 steps in 25-step increments, starting from a polarized (droplet) configuration. The top row is with relocation, the middle row is with telephoning, and the bottom row is with no mixing at all. Notice that relocation is more spatially disruptive than telephoning. Simulations were run with $p_a = 0.01$, mixing = 0.02, and uniform influence. Further evolution of the systems, including a comparison to the voter model, can be found in the appendix.}
\end{figure}

Our general finding is that relocation is more spatially disruptive than telephoning (see Figure \ref{consensus}). That is, compared to the case where there is no mixing at all, well-defined clusters (e.g., a droplet) will break down faster when there is relocation than when there is telephoning. Visual inspection suggests that part of the reason for this is related to clusters dissolving ``from the inside out.'' Relocation allows for extreme opinions of the other type to suddenly appear anywhere in a cluster. This is not the case for telephoning. Because agents retain their spatial locations during telephoning, an extreme opinion at a given site in the cluster can only become an extreme opinion of the other type by moving incrementally across the attitude spectrum. Consequently, even if an opinion at such a site moves towards the other end of the spectrum, there are many opportunities where this will be reversed by interactions with local neighbors. While the appearance of opposite extreme opinions can also be reversed in a cluster after relocation, it generally takes more local  interactions for this to occur. In short, relocation is more efficient at combating the effects of reinforcement than telephoning because it introduces more variation of opinion types within a cluster than does telephoning. 

\begin{figure}
\centering
	\begin{subfigure}[t]{0.3\textwidth}
	\includegraphics[width=\textwidth]{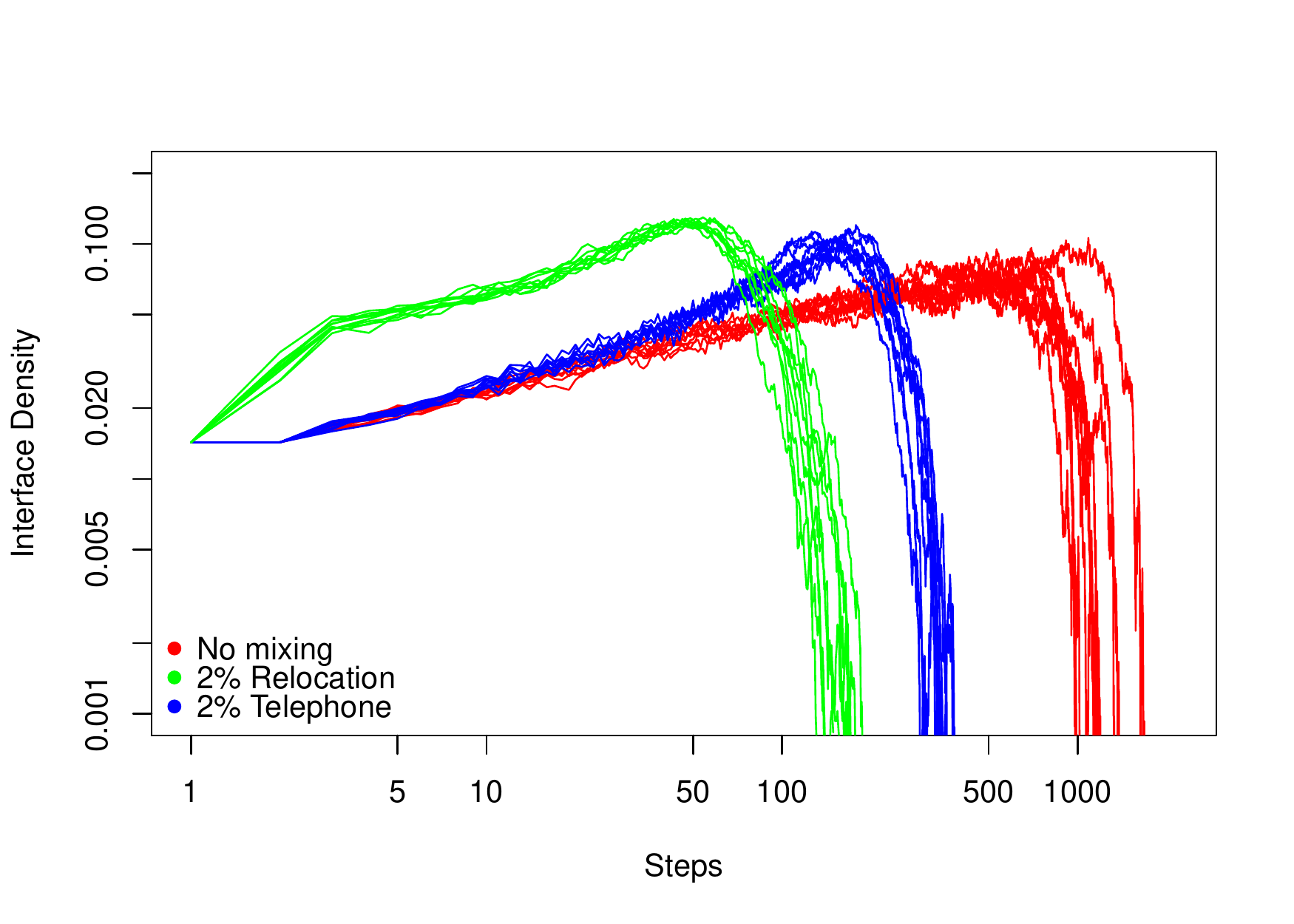}
\caption{Amp=0.001}
	\end{subfigure}
    	\begin{subfigure}[t]{0.3\textwidth}
	\includegraphics[width=\textwidth]{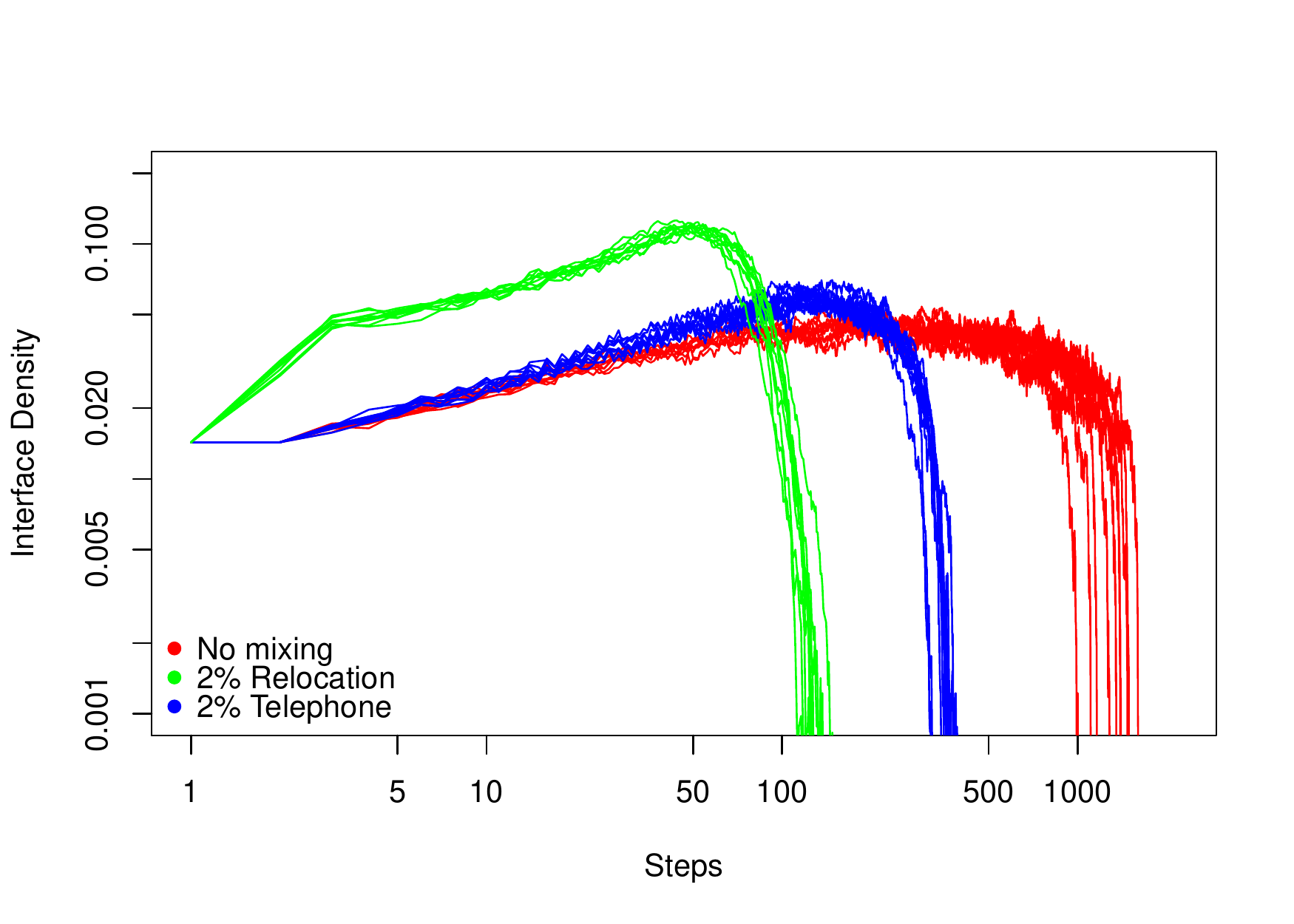}
\caption{Amp=0.01}
	\end{subfigure}
    	\begin{subfigure}[t]{0.3\textwidth}
	\includegraphics[width=\textwidth]{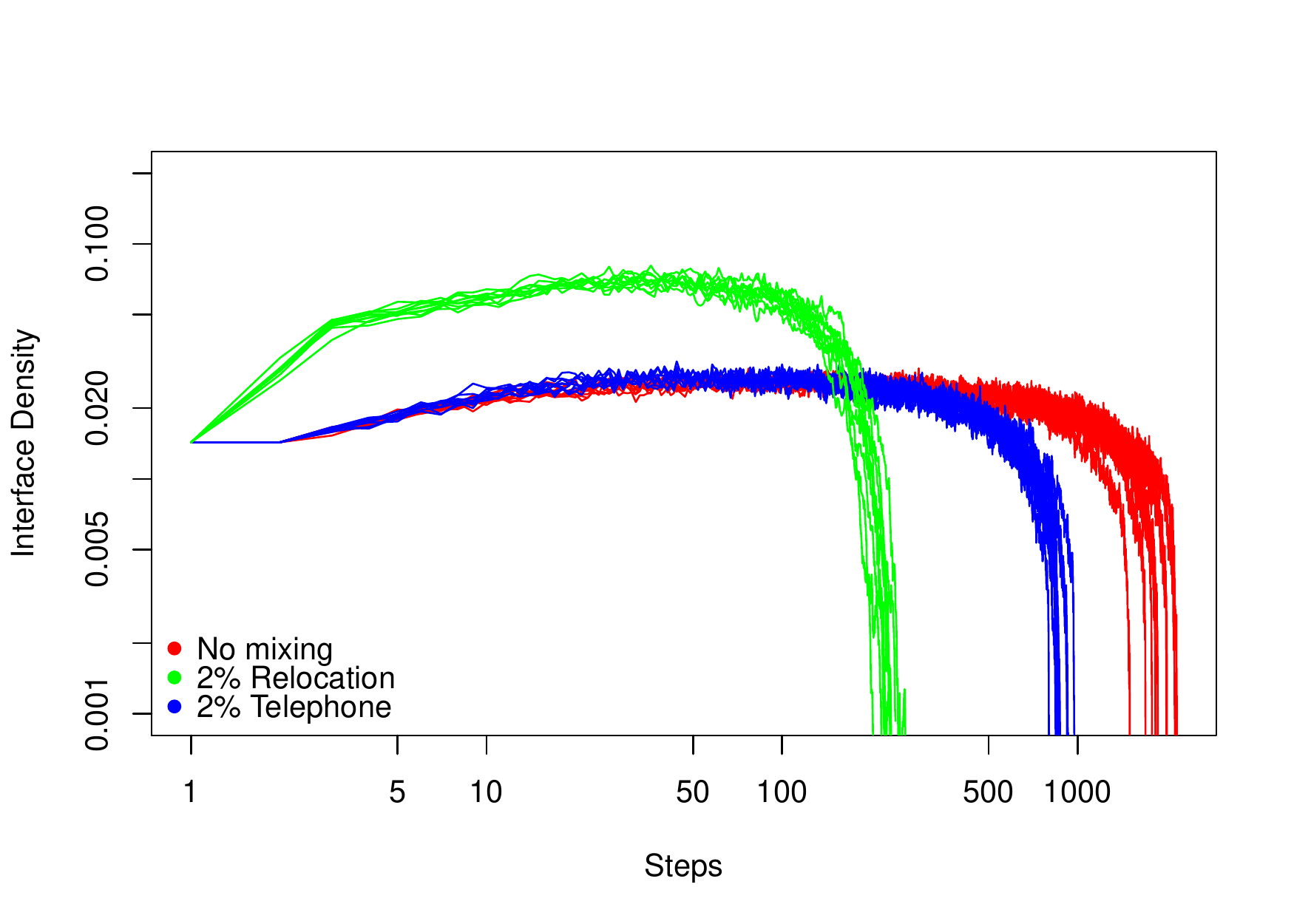}
\caption{Amp=0.01}
	\end{subfigure}
\caption{\label{interface_density} (Color online) Log-log plots of changes in interface density during droplet simulations (see Figure \ref{consensus}).  Each type of mixing scenario is shown, with three amplification levels. When interface density hits zero, consensus is reached. Note that, consistent with our visual inspection, relocation has consistently shorter consensus times than telephoning, and in turn telephoning has shorter consensus times than the case of no mixing. Moreover, relocation is more spatially disruptive, as indicated by the higher levels of interface density during the earlier time steps. }
\end{figure}

In addition to visual inspection, we analyzed simulations using interface density, \cite{sood2008voter,suchecki2005,dornic2001}. Lower values of interface density correspond to smoother and more well-formed boundaries (like the droplet) while higher values of interface density correspond with more noise.  Figure \ref{interface_density} shows that indeed, the different types of mixing differentially affect the spatial dynamics of the system, which in turn affect consensus times. Specifically, during droplet experiments, relocation produces the highest amount of interface density because it breaks up the droplet (as described in the paragraph above).

\subsection{Influence Functions}
\label{sec:influence-results}

The results discussed so far assumed uniform influence for local interactions; that is, each attitude has the same chance of being selected (given equal frequencies of the attitudes) because each attitude has the same amount of influence. We also considered four additional influence functions, two `extremist' (linear and quadratic) and two `centrist' (co-linear and co-quadratic). The linear and quadratic influence functions give strongly held opinions more influence; the co-linear and co-quadratic functions give moderately held opinions more influence. 

\begin{figure}
\centering
\includegraphics[width=0.9\textwidth]{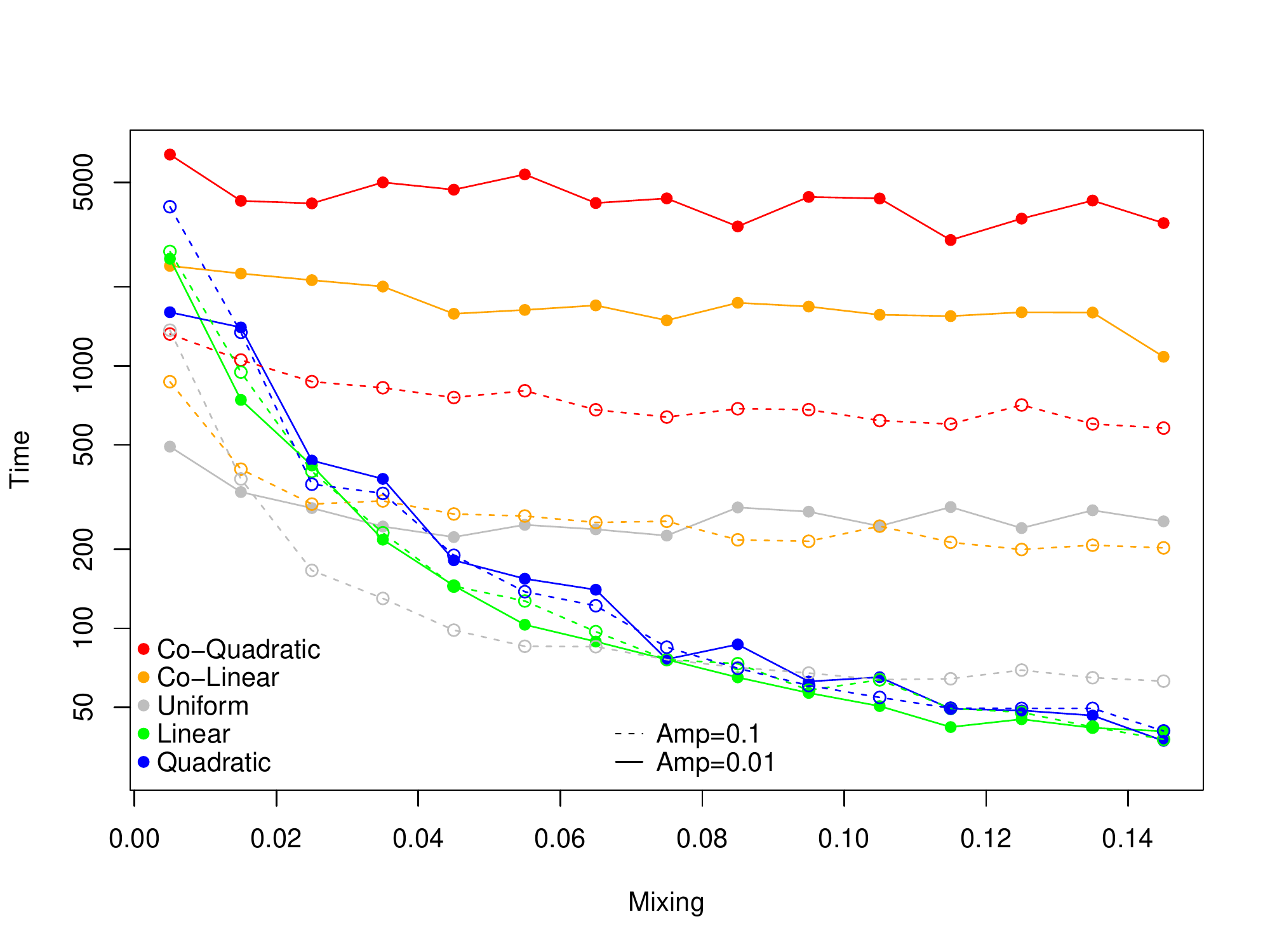}
\caption{\label{reloinfluence} (Color online) Mean consensus times (on a log scale) for various influence functions as a function of levels of mixing by relocation. (Telephoning results are not shown, but compare Figure \ref{timelogplot}).) Amplification and the Linear and Quadratic influence functions are processes that ``favor'' more entrenched opinions. The combination of these processes with increased mixing significantly decrease consensus times (see green and blue lines, and dashed grey line). Data are plotted for two levels of amplification, very low ($p_a=0.01$) and low ($p_a=0.1$).  The left-most data points correspond to mixing frequency 0.005. The case of no mixing is not shown. Each point is the mean of 50 simulations.}
\end{figure}

Figure \ref{reloinfluence} shows consensus times for the five different influence functions given low and very low amplification (amp=0.1 and amp=0.01, respectively) as a function of mixing by relocation (the case of telephoning is similar for sufficiently higher values; see Figures \ref{relativetime} and \ref{timelogplot}). For centrist influence functions, there is very little effect of mixing on time to consensus. For extremist influence functions, an increase in mixing speeds up time to consensus. The uniform influence case also sees an effect of mixing, most noticeably when amplification is higher. 

\section{Discussion}
\label{discussion}

Models allow us to investigate how separate psychological and sociological features could impact population-level phenomena related to opinion dynamics. 
On the psychological side, we can investigate how information presented to an individual is \textit{integrated} into their system of opinions or beliefs. Numerous biases have been studied, including biased assimilation \cite{lord1979biased,miller1993attitude,munro2002biased,taber2006motivated}, the ``myside" bias, or confirmation bias \cite{jonas2001confirmation,wason1960failure,wason1968reasoning}. The main idea is that one's initial opinion biases subsequent opinion updates, such that information consistent with one's opinion tends to be integrated while contrary information tends to be discounted. In this paper we investigated the effects of \textit{amplification}. Amplification shares similarities with the previously mentioned biases, but only focuses on how information could be integrated \textit{when it is consistent} with an individual's current opinion. In other words, amplification provides a bias that is active when there is agreement, not when there is disagreement. 

The sociological side is concerned with how \textit{interaction partners} are ``chosen." There are two broad categories here.  \textit{Unstructured} interactions happen when individuals meet randomly, while \textit{structured} interactions happen when there is something that systematically determines which individuals meet. One such example is \textit{homophily}, where interactions between like-minded individuals are more frequent  \cite{mcpherson2001birds,smith2014social,halberstam2016homophily}. We explored structured interactions with our \textit{influence} functions. Centrist influence functions bias interactions toward more centered opinions, while extremist influence functions bias interactions toward more entrenched opinions. We also explored the relationship between structured and unstructured interactions by introducing two types of \textit{mixing} mechanisms: relocation, where some fraction of the population changes their location, and telephoning, where some fraction of the population temporarily interacts with agents outside their local neighborhood. Without mixing, interactions are as structured as they can be in our models, i.e., agents only interact with others in their local neighborhood. As we increase mixing, interactions become less structured, i..e, agents increasingly interact with randomly selected agents in the population.

Centrist and extremist influence functions modulate our results slightly (see Figure \ref{reloinfluence}). Simulations suggest that these findings continue to hold qualitatively when $L=3$ or $L=4$, though on slightly longer timescales (results not shown). When $L=1$ we simply have the voter model \cite{baumgaertner2016opinion}.

Our results echo previous findings in the opinion dynamics literature that say polarization is the result of specific psychological and sociological processes that are combined. For example, Dandekar et al.\ \cite{dandekar2013biased} show that \textit{the combination} of biased assimilation and homophily produces polarization. Homophily on its own does not. For example, if the psychological process is DeGroot-like, where individuals update their opinion as a weighted average of their current opinion and that of their neighbors, then polarization does not emerge, \textit{even if} the population has a high degree of homophily. For homophily to produce polarization, it needs to be combined with biased assimilation (or something like it). We have a similar result. In our spatial model, clusters of opinions form from an initial random configuration of attitudes, i.e., we get homophily. However, in the case of uniform influence and no amplification, or the case of centrist influence functions with sufficiently low levels of amplification, entrenched opinions disappear over time, eventually converging to the special case of only two centrist attitude states, ${\cal A} = \{-1,1\}$. (Here our spatial model behaves like the (discrete time) voter model \cite{holley1975ergodic,clifford1973model,durrett1988lecture,liggett}.) In order for our spatial model to produce polarization, amplification (our analog of biased assimilation) must be sufficiently high.

In addition, we show that amplification is not enough to produce polarization by itself. As all our well-mixed models illustrate, increasing amplification decreases the time it takes for a population to reach consensus. It is only when a sufficient amount of spatial structure is maintained by keeping mixing low, in addition to a small amount of amplification, that we obtain polarization. So again, it is \textit{the combination} of structured interactions with opinion amplification that produces polarization. To be clear, spatial structure itself is not what generates clustering (or homophily), but the opinion formation process that happens on the spatial structure.  Clustering in turn produces polarization when adding even a small amount of amplification.

Bounded confidence models have also been used to study polarization
\cite{Hegselmann02opiniondynamics,deffuant2002can,dandekar2013biased,salzarulo2006,deffuant2000mixing,weisbuch2003interacting}. In these models, agents that become sufficiently dissimilar with respect to their opinions cease to influence one another; agents can become ``closed-minded.'' In our model, no matter how entrenched an opinion becomes, that agent's attitude can still be changed through the influence of other agents holding opposing opinions. 
Moreover, our influence functions are symmetric, which means the direction that agents feel pulled is not determined by the influence functions themselves, but rather by the frequencies of opinions (both locally and globally, where global frequencies of opinions will dominate as levels of mixing are increased). In brief, we show how polarization can arise, not by agents becoming ``close-minded,'' but by the combination of some psychological bias and structured interactions. 

It is worth pointing out that some opinion dynamics models also focus on how opinion diversity can be maintained or generated. The recent ISC (influence, susceptibility, and conformity) model is a notable example \cite{duggins2017}. In the ISC model, diversity of opinions can be maintained because individuals end up being pulled towards center and extremism simultaneously in a population that balances heterogeneous intolerance, susceptibility, and conformity (we leave aside the details of these concepts and refer the reader to the original article). We have two ways of generating or maintaining opinion diversity. In \cite{baumgaertner2016opinion} we argued that, in the fully spatial case, diversity of opinion or attitudes can be maintained by counter-balancing amplification with co-influence functions. The second way of sustaining opinion diversity is with a very small amount of mixing: no mixing and some amplification produces polarization, and a small to large amount of mixing hastens consensus, but in between these cases it is possible to maintain a roughly uniform distribution of opinions for some time (in the limit, however, consensus is eventually reached, but on such a long timescale that is not of interest).  How much mixing is required to generate and sustain diversity will be less than it takes to hasten consensus, but will still depend on the frequencies of opinions, the level of amplification and which influence function is used. In any case, as the ISC model assumes that agents have fixed locations, mixing is an interesting point of difference.

While diversity of opinion is an important phenomenon to capture, our primary focus is on the impact of levels of mixing on reaching consensus. One of the most striking results in this study is that the very conditions (e.g., amplification) which lead to polarization and stagnation in the strictly spatial model produce a tipping point with rapid consensus in the presence of sufficient mixing. Moreover, this effect is observed across population sizes. This tipping point occurs when the frequency distribution of attitudes in the attitude spectrum becomes asymmetric enough to rapidly pull the rest of the individuals over to the same opinion. The amount of mixing it takes to go from very long consensus times to very short ones depends on the type of mixing, the amount of amplification, and the influence function. As a rule of thumb, very low levels of amplification (e.g., 1\%) tend to produce similar consensus times across different levels of mixing. For higher levels of amplification (e.g., 10\%), however, higher levels of mixing (e.g., 10\%) will dramatically \textit{decrease} consensus time (relative to the low amplification case), while low levels of mixing (e.g., 1\%) dramatically \textit{increase} consensus time. 

It is interesting to compare these results to other opinion dynamic models with mixing. For example, Castellano et al. \cite{castellano2003} consider a voter model on a 1-dimensional torus, with and without the addition of small-world edges. As in our model, the addition of long-range interactions produces a tipping point, with a period of diversity followed by a rapid transition to consensus. In their case, however, the time to consensus depends strongly on system size, $N$. This is also true of the 1-dimensional voter model, for which mean consensus time grows with $N$.  Of course, the case of infinite system size and complete mixing results in the mean-field ODE, $dx/dt = (1-x)x - x(1-x)=0$, that has the fraction of 1's never changing. Our ODE, by contrast, has the same tipping point behavior we see in the spatial model with partial mixing. This agrees with the fact that consensus time in our model is roughly independent of $N$ when $N$ is large (Figure \ref{systemsize}).

Care must be taken to consider appropriate regions of the parameter space in our entrenchment model; after all, real populations do not tend to reach consensus rapidly (if at all). The regions of parameter space that make intuitive sense produce patterns reflected in real populations. For example, low levels of mixing allow clusters of opinions to emerge, which corresponds to homophily in real populations. These clusters in turn increase time to consensus. Moreover, if amplification is low but still non-zero, clusters will ultimately lead to polarization, causing deadlock. If, however, we introduce some mixing, then clusters will undergo some changes.  Our model thereby makes an empirical prediction. Suppose we have two sufficiently large groups of otherwise similar individuals discussing some matter that requires group consensus: Group 1 is highly structured in their interactions, while the interactions in Group 2 are random (approximating our mixing scenarios). Our models suggest that groups of type 2 will tend to reach consensus more quickly than groups of type 1, and that the difference in time will be greater for groups of individuals with higher levels of biased assimilation or confirmation bias (approximating our levels of amplification).

\section{Conclusion}

We considered several versions of our general entrenchment model of opinion dynamics: the fully spatial model, the telephoning model, and the relocation model. The behavior of the telephoning and relocation models diverge for a small amount of mixing and come together as mixing increases, where sufficiently high levels of mixing can be approximated by an ODE. Real populations are somewhere in between, leaning towards less mixing: sometimes people move to new communities, sometimes people have interactions outside their normal contacts, but most interactions are with the same people from a relatively small group. 

We compared the effects of these two modes of mixing on the dynamics of opinion formation.
In previous work, we analyzed the effects of opinion amplification in a population where individuals interacted only locally, and we found that amplification produces clusters of opinion and polarization towards more extreme opinions resulting in long-term deadlock. There we compared our model to other existing models of opinion dynamics, including bounded confidence models and models that explore mechanisms that produce polarization \cite{friedkin2015problem, deffuant2002can,Hegselmann02opiniondynamics,deffuant2000mixing,flache2011small,dandekar2013biased}. Our findings show that the effect of polarization by amplification, which leads to deadlock or at least increased time to consensus, is reversed in a well-mixed system; an increase in amplification \textit{decreases} the time to consensus. The transition from deadlock to consensus as we move from a purely local to mixed population depends on the type of mixing. 

Our findings suggest that mixing by relocation will reverse deadlock faster than mixing by telephoning. Where this reversal happens and how much faster it occurs depends on the level of amplification. As amplification probability approaches zero, the difference between relocation and telephoning is negligible. However, as amplification is increased, even just a small amount, the difference between relocation and telephoning becomes significant. The combination of relocation and amplification dramatically decreases the time to consensus, quickly approaching the behavior of the ODE system. On the other hand, it takes much more telephoning (in combination with amplification) to approach the same consensus time behavior.

\appendix

\section{Derivation of the Mean-Field ODEs}
\label{app:ODEderivation}

We begin by focussing on the ODE for the left-most extreme opinion frequency, $L_2$.  This population decreases when $a=-2$ individuals become $a=-1$ individuals, and increases when $a=-1$ individuals become $a=-2$ individuals.  Since attitudes can only move one step at a time, these are the only possible loss and gain interactions.  

The loss interactions are:
\begin{description}
\item[L1] an $a=-2$ focal individual has a no-amplification interaction with an $a=-1$ individual
\item[L2] an $a=-2$ focal individual has an interaction with an $a=1$ individual
\item[L3] an $a=-2$ focal individual has an interaction with a an $a=2$ individual
\end{description}
Notice that all of the losses to the $-2$ population come from interactions of the $-2$ population with other attitudes.  The remaining $-2$ interactions are steady-state interactions that result in no change in $L_2$.  These interactions are
\begin{description}
\item[SS1] an $a=-2$ focal individual interacts with another $a=-2$ individual
\item[SS2] an $a=-2$ focal individual has an amplification interaction with an $a=-1$ individual
\end{description}
We can thus either count up the three loss interaction types ({\textbf{Li}} interactions), or subtract the two steady-state interaction types ({\textbf{SSi}}) from the total number of -2 interactions, which works out to be simply the frequency of -2 individuals, or $L_2$.  More formally, we have
\beq
{\textbf{L1}}+{\textbf{L2}}+{\textbf{L3}} = L_2-({\textbf{SS1}}+{\textbf{SS2}})
\label{eq:ss-interactions}
\eeq
The gain interactions are
\begin{description}
\item[G1] an $a=-1$ individual has an interaction with with an $a=-2$ individual
\item[G2] an $a=-1$ individual has an amplification interaction with another $a=-1$ individual
\end{description}
Note that {\textbf{SS2}} is different from {\textbf{G1}}.  In the {\textbf{SS2}} interaction the focal individual has attitude -2, while in {\textbf{G1}} the focal individual has attitude -1.  The interactions and their rates are summarized in Table~\ref{tbl:interactions-to-ODE}.
\begin{table}
\begin{tabular}{c|cccc}
\, interaction \, & focal & interaction & loss $(-)$ & rate \\
label & \, attitude \, & partner & or gain $(+)$ & \\
& & attitude & & \\
\hline
L1 & -2 & -1 no amplification & - & $(1-p_a) L_2 L_1$ \\ 
L2 & -2 & +1 & - & $L_2 R_1$ \\
L3 & -2 & +2 & - & $L_2 R_2$ \\
SS1 & -2 & -2 & 0 & $L_2 L_2$ \\
SS2 & -2 & -1 with amplification & 0 & $p_a L_2 L_1$ \\
G1 & -1 & -1 with amplification & + & $p_a L_1 L_1$ \\
G2 & -1 & -2 & + & $L_1 L_2$ \\
\end{tabular}
\caption{Table showing the interactions that increase, decrease, or keep steady the $L_2$ portion of the population.  Increases are shown as gains (+), decreases as losses (-), and steady-state interactions as neither (0).  The interaction label is used to match these interactions with the terms in the ODE~\eqref{eq:ODE-2_labeled}.}
\label{tbl:interactions-to-ODE}
\end{table}

We can thus write the ODE for $\dot{L_2}$ as
\begin{align}
\frac{dL_2}{dt} & = ({\textbf{G1}} + {\textbf{G2}}) - ({\textbf{L1}} + {\textbf{L2}} + {\textbf{L3}}) \nonumber \\ 
& = ({\textbf{G1}} + {\textbf{G2}}) - (L_2 - ({\textbf{SS1}} + {\textbf{SS2}})) \nonumber \\
& = (L_1 L_2 + p_a L_1^2) - (1-(L_2^2 + L_2 L_1)) \nonumber \\
& = L_1(L_2 + p_a L_1) - L_2 (1 - L_2 -p_a L_1)
\label{eq:ODE-2_labeled}
\end{align}
The ODE~\eqref{eq:ODE-2_labeled} is the same as~\eqref{ODE-1}.

The other ODEs in~\eqref{eq:ODEsimplified} are built in an analogous fashion.  The $\dot{R}_2$ equation is symmetric with the $\dot{L}_2$ equation.  The $\dot{L}_1$ and $\dot{R}_1$ equations are also symmetric with each other, and contain more terms since these subpopulations can be increased from two other subpopulations, rather than just one (see Figure~\ref{fig:ode-transitions}). 


\section{Stable Manifold of the Centering Model}
\label{app:stable-manifold}

When $p_a=0$ the centering model~\eqref{eq:ODE_2D_centering} becomes 
\beqsub
\beqa
\dot{x} & = & -y^2 + (x+1)y + (2x-1)x, \\
\dot{y} & = &y^2 + (x-1)y.
\eeqa
\label{eq:ODE_2D_centering_eps0}
\eeqsub
Define the Lyapunov function $L = 1/2-x-y$.
Then, taking the derivative of $L$ in the flow field defined by~\eqref{eq:ODE_2D_centering_eps0} we obtain
\[
\dot{L} = -\dot{x}-\dot{y} = -2xy-2x^2+x.
\]
The line $L=0$ is an invariant set of the dynamical system if
\[
\dot{L}=0 \Leftrightarrow 2xy+2x^2-x=0 \Leftrightarrow x=0 \text{ or } x+y=\frac{1}{2}.
\]
With the second condition, we recover the line $L=0$.
We conclude that the solution curve emanating from any point on the line $L=0$ (or $x+y=1/2$) remains on that line.  The direction of flow for solutions on that line is given by $\dot{x}$ and $\dot{y}$ using $x=1/2-y$ and $y=1/2-x$.  We obtain
\beqa
\dot{y} & = & y^2 + \left(\frac{1}{2}-y-1\right)y = -\,\frac{1}{2}y, \nonumber \\
\dot{x} & = & -\left(\frac{1}{2}-x\right)^2 + (x+1)\left(\frac{1}{2}-x\right) + (2x-1)x = \frac{1}{2}\left(\frac{1}{2}-x\right). \nonumber
\eeqa
Thus, all initial points on $L$ that satisfy $y>0$ and $x<1/2$ yield solutions that flow in the direction of decreasing $y$ ($\dot{y}<0$) and increasing $x$ ($\dot{x}>0$), while initial points on $L$ that satisfy $y<0$ and $x>1/2$ flow in the opposite direction.  We conclude that $L$ is the stable manifold for the steady state (0.5,0).

When $p_a>0$, the stable manifold is no longer the line $L$, but numerical simulations indicate that the new stable manifold is close to the original one.

\section{Droplet Experiments and Surface Tension}

\begin{figure}
\centering
\includegraphics[width=.99\textwidth]{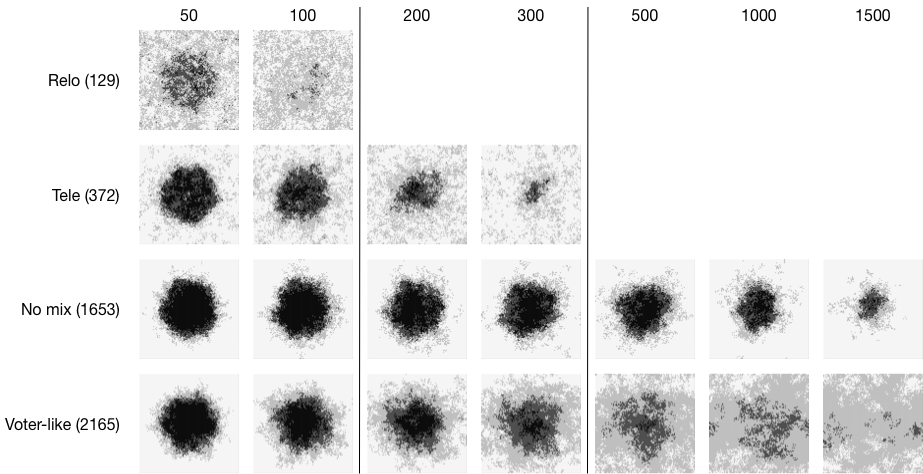}
\caption{\label{droplet} Evolution of attitudes over 1500 time steps in varying increments, starting from a polarized (droplet) configuration. The number of plots in each row can vary because the time to consensus (in parentheses in the left margin) is different in each case. The first row is with relocation and the second with telephoning, both at 0.02. The third row shows the case without mixing. Each of these three scenarios has $p_a = 0.01$.  The bottom row is the case without amplification, which is similar to the voter model. Notice that without amplification (row four) there is a lack of surface tension and the droplet diffuses, in contrast to the no mixing case (row three). A similar diffusing effect is achieved by relocation (row one), but the level of mixing by telephoning (row two) is not sufficiently disruptive and some surface tension persists.}
\end{figure}

\newpage
\bibliography{relocation.bib}

\end{document}